# The Future of Child Development in the AI Era:
## Cross-Disciplinary Perspectives Between AI and Child Development Experts

Neugnot-Cerioli, Mathilde
Muss Laurenty, Olga
mathilde@everyone.ai

## ABSTRACT

This report explores the potential implications of rapidly integrating Artificial Intelligence (AI) applications into children's environments. The introduction of AI in our daily lives necessitates scrutiny considering the significant role of the environment in shaping cognition, socio-emotional skills, and behaviors, especially during the first 25 years of cerebral development. As AI becomes prevalent in educational and leisure activities, it will significantly modify the experiences of children and adolescents, presenting both challenges and opportunities for their developmental trajectories. This analysis was informed by consulting with 15 experts from pertinent disciplines (AI, product development, child development, and neurosciences), along with a comprehensive review of scientific literature on children development and child-technology interactions. Overall, AI experts anticipate that AI will transform leisure activities, revolutionize education, and redefine human-machine interactions. While AI offers substantial benefits in fostering interactive engagement for example, it also poses risks that require careful considerations, especially during sensitive developmental periods. The report advocates for proactive international collaboration across multiple disciplines and increased research into how technological innovations affect child development. Such efforts are crucial for designing a sustainable and ethical future for the next generation through specific child-centered regulations, and helping to educate all potential stakeholders, (regulators, developers, parents and educators, children) about responsible AI use and its potential impacts on child development.

## 1. INTRODUCTION

The concerns regarding Artificial Intelligence's (AI) impact on children's lives invoke a timeless debate about generational perspectives on technology and progress, reminiscent of Horace's observation from 20 BC that each generation views its "progeny yet more corrupt". While it's crucial to question whether anxieties about AI's influence on children in their environment are merely a new iteration of this phenomenon, it is equally important to acknowledge the unprecedented speed at which these changes are occurring. Further, history shows that humans have consistently regulated and legislated the use of new technologies to ensure safety, tapping into their full potential while leveraging the risks, like what appears necessary in the field of AI. Knowledge about cognitive and socio-emotional child development, along with early findings on the impact of technology on children and adolescents, reveals that these interactions are complex and multifaceted. This warrants careful consideration and study to develop appropriate safeguards. Therefore, although the sentiment of generational apprehension towards new technology is not new, the unique challenges and opportunities presented by AI in children's lives call for a nuanced, evidence-based approach.

The development of a child's brain is the result of both genetic and environmental factors. From the prenatal stage through early adulthood around 25 years old, the brain undergoes significant growth and transformation, with environmental stimuli playing a pivotal role in shaping neural pathways and cognitive functions. Early childhood and adolescence are marked by sensitive periods where plasticity makes their brain particularly receptive to external influences. The traditional elements of a child's environment – familial interactions, educational settings, and social experiences – have been extensively studied in terms of their impacts on developmental outcomes. However, the rapid integration of AI into this environment introduces a novel and complex variable into the equation.

The main goal of this report is to create a collaborative framework that brings together experts in technology, cognitive neurosciences and child development. The purpose is to anticipate the potential impact of AI applications on cognitive development and overall well-being in children. To achieve this, the paper employs two key strategies:

1) **Consult with experts in AI, product development and child development:** The goal is to provide a general understanding of how AI is likely to be applied in children's environments and to anticipate some of the potential benefits and challenges it represents.

2) **Conduct an extensive review of the scientific literature:** The focus of this research is on child development, child-technology interactions, and regulations, to gain insights on the implications of growing up in a generalized AI-infused environment.

The contributors' intent is not to create an exhaustive list of all potential implications, but to start a collaborative effort to incorporate AI into children's lives in a way that is responsible, beneficial, and informed by a wide range of interdisciplinary research insights.

## 1.1. Current context of artificial intelligence/Machine Learning

Though the notion of AI/Machine Learning (ML) can be traced back to the 1940s, the advent of Generative AI (GenAI) marked a significant turning point with the public release of ChatGPT-3.5 by OpenAI in November 2022 (Cao et al., 2023). This release saw rapid adoption, gaining 1 million users within its first month and now, nearly two years later, boasting over 180 million monthly active users. Following ChatGPT-3.5, other tech companies released their own large language models (LLMs), such as Anthropic's Claude, Google's Gemini, Gemma (and other models) Meta's Llama, and others like Midjourney and Mistral. Consequently, AI has received increasing media attention worldwide, raising public awareness of AI's capabilities.

AI was already embedded in various sectors, including data mining, industrial robotics, logistics, business intelligence, banking, medical diagnostics, recommendation systems, and search engines (Delipetrev et al., 2020). However, it has become more tangible for the public, since it now allows for more direct, conscious interactions. AI technologies were already present in various aspects of daily life, often without users being fully aware of their presence. In many cases, AI operates behind the scenes, enhancing the functionality and user experience of numerous products and services.

Historically, AI "artificial intelligence" or "intelligence demonstrated by machines" is often described as artificially developed system that exhibits understanding, problem solving, among other tasks by understanding how humans think and simulating human intelligence (Delipetrev et al., 2020). The large language models (LLMs) we see today are built on a series of advancements in mathematics and engineering, increased datasets—largely thanks to the internet—and enhanced computational power (LeCun et al., 2015). Perhaps more surprisingly, its advancement has been tightly influenced by breakthroughs in the fields of cognitive psychology and neuroscience. Indeed, artificial intelligence has largely been developed to mimic human intelligence, though focusing primarily on its cognitive aspects, such as data processing, pattern recognition, and decision-making, rather than emotions and social skills, which are critical elements of human intelligence (Mitchell, 2020).

The 1980s to 1990s saw the development of **machine learning**, a subset of AI that, unlike simple algorithms, allows computers to learn from data and improve their performance over time. It was significantly influenced by cognitive science, with reinforcement learning—a fundamental machine learning technique—based on principles from behavioral psychology and conditioning, where rewards increase the likelihood of behaviors being repeated (Mitchell, 2020).

Starting in the 1990s, AI models became more refined with the development of **deep learning**, enabling more complex processing tasks. On one hand, **convolutional neural networks (CNNs)**, inspired by the discovery in neuroscience of how the eye detects edges, textures, and complex patterns, led to advances in visual, auditory, and pattern recognition (Mitchell, 2020). On the other hand, **recurrent neural networks (RNNs)** allow for handling sequential data, similar to how the brain processes sequences of information, like language or pattern recognition, learning from past information to predict future events (Mitchell, 2020).

Starting in the late 2010s, the development of **Generative AI (GenAI)** marked a transformation from previous models (Bengesi et al., 2023). GenAI regroups models that can generate new content, such as text, images, music, and code, based on learned patterns from training data. GenAI models, particularly those using the Transformer architecture like GPT (Generative Pre-trained Transformers), undergo extensive pre-training on large datasets, allowing them to analyze and generate human-like content. This evolution was significantly influenced by advancements in deep learning and neuroscience. For instance, Transformer models mimic human attention mechanisms, reflecting how the brain focuses on important stimuli while processing information (Vaswani et al., 2017). These models are inspired by human creativity, where the generative process mirrors how humans imagine and create new ideas, making them effective in text completion, translation, and creative writing.

AI technology is advancing at an unprecedented pace, with its exponential growth accelerating significantly since around 2010; the computational capacity for training began to double every 6 months, going faster than Moore's Law, which predicted a doubling every 20 months (measured in FLOPs, a unit of operation in an artificial neural network, (Sevilla et al., 2022). This rapid acceleration in technological innovation leads to changes in our environment happening at a faster pace than our ability to study its potential impact on human's behaviors.

Looking ahead, AI is poised to continue its rapid trajectory deeply transforming our lives and industries, with ongoing research in areas such as reinforcement learning, quantum computing, and AI ethics (Castelvecchi, Davide, 2024). As AI technologies become increasingly integrated into various aspects of daily life and industry, they offer tremendous societal benefits while also raising important broader ethical questions about privacy, security, and the future of work. The ongoing evolution of AI will undoubtedly shape the technological landscape for decades to come, and become further embedded into educational, recreational, and social domains. New technologies, such as Immersive Virtual Reality (IVR), which regroup both Virtual Reality (VR) and Augmented Reality (AR) are being widely used and adopted (Meta, 2023). Also, monitoring through biometrics and physiological data is rapidly expanding, offering both promises and notable challenges, specifically for Physiological & Physical AI Monitoring, and Child Development AI Monitoring. In this context, it is imperative to consider how these changes might affect developmental trajectories.

In this report, to allow for more concise language the term "AI" is used to refer to both the technology itself and the applications, objects, and use in which it is integrated.

### 1.2. Child Development

Skills, thoughts, and behaviors are shaped and determined by brain functions. Individual differences arise from recursive and reciprocal child-environment interactions, while genetics set up differential susceptibility to environmental conditions. Genetics provides the blueprint for the development of the nervous system and sets its timeline, while the environment influences how the brain develops. At the intersection of genetics and environmental factors lie sensitive periods, when the brain is primed to develop certain skills, and when missed opportunities can have lasting consequences (Banich & Compton, 2023).

The brain contains an average of 80 billion neurons (Azevedo et al., 2009) that convey information: they first receive then relay information to other neurons, and those connections are called synapses. The development of those cells and the way they transmit information effectively happens through four main physiological processes: neuronal proliferation, synaptogenesis, pruning and myelination. This begins with **neuronal proliferation**, which occurs during early embryonic stages and continues for a period after birth (Silbereis et al., 2016). This sets the stage for **synaptogenesis**, the formation of neural connections or synapses, which is particularly crucial in early childhood and peaks around the age of three (Huttenlocher, 2009). Genetic blueprints determine the timing and location of these processes. For instance, synapse formation in the auditory cortex reaches its peak at three months, whereas in the prefrontal cortex, it continues until the child is about three years old. Post-synaptogenesis, the brain undergoes further sculpting through pruning and myelination. **Pruning** eliminates unnecessary or underused synapses, streamlining brain functions for optimal performance (Huttenlocher, 1990; Huttenlocher & Dabholkar, 1997). **Myelination**, the insulation of nerve fibers with a protective myelin sheath, enhances the efficiency of neural networks and extends from fetal development into early adulthood (Simmonds et al., 2014).

To underscore the significance of the environment on development, consider the Abecedarian Project of the early 1970s, which divided children from low socioeconomic backgrounds into two groups - one receiving enhanced nutrition and a stimulating environment, while the other did not. The findings were striking: the enriched group exhibited significantly improved cerebral connectivity, translating to higher IQ scores, and higher well-being and success into adulthood. (Campbell et al., 2008, 2012; Muennig et al., 2011). This study vividly illustrates how early environmental factors can profoundly influence developmental outcomes.

The interaction of children and adolescents with AI-infused technology in the future could potentially modulate and influence overall development. Indeed, the development of skills is influenced by two crucial factors: the interplay between environmental experiences, which are categorized as either expectant or environment-dependent (Bruer & Greenough, 2016), and sensitive periods, during which the brain is especially receptive to developing certain skills.

First, "**Experience-Expectant System**" refers to neural systems that are shaped by experiences universally present in typical development (Bruer & Greenough, 2016). The formation of these neural features depends on external information that is essential for development but not specified in the genetic blueprint. Abilities reliant on this system develop normally given typical environmental exposure; however, lack of such experience can result in significant disabilities. For instance, infants' visual systems are primed to develop in response to visual stimuli. If deprived of visual input during the first few months of life due to congenital cataracts, this can lead to lasting visual processing deficits, even if sight is later restored through intervention (Lewis & Maurer, 2005).

The second category of environmental experiences is termed "**Experience-Dependent Systems**." These systems are uniquely shaped by individual encounters and interactions within one's environment and are likely to be significantly influenced by the pervasiveness of AI in the future. Abilities such as playing an instrument, speaking multiple languages, or excelling in sports are good examples of this type of neural systems (Banich & Compton, 2023). Exposure to such experiences in the environment modifies brain connectivity and activation, which in turn correlates with behavioral changes.

An additional important concept in child development is how changes in the environment, such as the integration of AI, might reshape learning and interactions during **sensitive periods of development.** Sensitive periods refer to specific times during development when an organism is particularly receptive to certain external stimuli, enabling learning to occur more rapidly and deeply. After these periods, acquiring certain skills may not be as efficient or effective (Banich & Compton, 2023). For example, research on children raised in orphanages under conditions of social deprivation has shown alterations in white matter (Sheridan et al., 2012) associated with lower IQ scores for those not placed in enriched environments before the age of 2, compared to those who were (Windsor et al., 2013). Likewise, learning a second language becomes significantly more difficult after the age of 17, with achieving native-like fluency unlikely (Hartshorne et al., 2018). The development and quality of AI applications hold both the potential to support developmental milestones and the risk of hindering them if they detract from crucial environmental interactions. This concept aligns with displacement theory, which posits that time young children spend on screens or adolescents on social media could supplant activities more conducive to development. Recognizing this dynamic is essential for guiding the design and use of technology in ways that foster rather than obstruct a child's comprehensive development (Boone et al., 2007; Putnick et al., 2023; Roberts et al., 1993; Schwarzer et al., 2022).

While the first years of a child's life are widely recognized as critical for development, as evidenced by initiatives like First 5 California (*First 5 California*, n.d., p. 5), the significant cerebral development occurring during adolescence is often overlooked. Scientific findings highlight this crucial period for brain development, typically between ages 12 to 20, characterized by pruning and myelination in key cortical regions. A notable aspect is the developmental mismatch between the rapidly maturing limbic system, associated with emotional responses and reward processing, and the more slowly developing prefrontal cortex, responsible for decision-making, risk assessment, and impulse control (Casey et al., 2008; Steinberg et al., 2018). This leads to adolescents often making riskier choices when rewards are involved despite their ability to reason logically in other situations (Shulman et al., 2016; Steinberg et al., 2018). The maturation of the limbic system coincides with the emergence of various psychological distresses, including an increased risk for depression (Altemus et al., 2014). Amid these changes, adolescents face a complex social environment, challenging their understanding and navigation of intricate social dynamics, emotional regulation, and decision-making (Blakemore, 2012). Adolescents do not yet possess the adult capacity to navigate these complexities, as evidenced by differing brain activity when interpreting others' mental states (Andrews et al., 2021). Further, since technology is moving so quickly, parents are often not equipped to guide them in this fast-paced digital environment. The convergence of a maturing brain, a developmental imbalance between cognitive and emotional processing systems, and escalating social demands highlights the vulnerability of adolescence. This critical phase underscores the need for a deep understanding of brain development to foster healthy growth and maturation, meriting special attention.

### 1.3. AI in Children's Environment

From interactive learning tools to algorithm-driven content on digital platforms, AI has become a pivotal component of children's daily lives. This evolution builds upon earlier technological shifts introduced by television, the internet, smartphones, tablets, and video games. The current digital landscape represents an important environment shift, with potential impacts on children's cognitive and emotional development that are not yet fully understood. Engagement with digital media has seen a marked increase across generations, with an average 17% rise since the COVID-19 pandemic began in early 2020; the typical American teenager now spends an average of over 8 hours a day on screen, not including the time spent using screens for school or homework (Rideout, V., Peebles, A., Mann, S., & Robb, M. B., 2022). Furthermore, the initial age of exposure to technology has dramatically decreased from 4 years in the 1970s to as young as 4 months old today (Radesky & Christakis, 2016). AI is now integrated in a large part of children's digital environments, such as in content curation in streaming services and social media, and gameplay in video games. With the advent of more sophisticated generative AI, two-thirds of adolescents reported using such technology in 2023 (FOSI, 2023).

Research into the effects of technology on children is intricate, influenced by the nature of screen activities, the age and environment of the child, and their unique cognitive profiles. Despite these complexities, there is some consensus on impacts of screens on several notable areas (Oswald et al., 2020). First, high screen use and limited outdoor activity are linked to rising myopia rates among children, particularly in Asia where up to 73% of South Korean teens are affected (Grzybowski et al., 2020; Jones-Jordan et al., 2012). Second, screen usage has been linked to disrupted sleep patterns, reducing both the quantity and quality of sleep, particularly when used before bedtime. This disruption can significantly affect cognition and mental health (Cheung et al., 2017; Khan, 2023; Magee et al., 2014; Riesch et al., 2019). Third, prolonged screen time contributes to increased sedentary behavior, raising the risk of cardiovascular issues later in life (Grøntved et al., 2014; Lona et al., 2021). Fourth, the impact of screens varies among different populations, with both more adverse as well as positive outcomes observed in individuals with neurodivergence (Gwynette et al., 2018; Weiss et al., 2011). Lastly, technoference, which relates to how technology interferes in human relationships is reshaping parent-child interactions, and constitutes a growing area of study as parents themselves express conflict over their screen usage around their children (Anderson et al., 2017; Gergen, 2002; Kildare & Middlemiss, 2017; Misra et al., 2016; Myruski et al., 2018; Pempek et al., 2014; Przybylski &

Weinstein, 2013; Radesky et al., 2014; Radesky & Christakis, 2016; Stockdale et al., 2020). Technology is influencing how children and adolescents relate to themselves and their peers, which can be negative in the case of cyberbullying for instance (Marín-López et al., 2020).

## 1.4. Current Regulations for Ethical AI

There is a growing consensus among regulators, policymakers, and AI researchers on the urgent need to properly regulate and create safeguards for the development and deployment of AI solutions. This consensus stems from multiple critical ethical considerations.

Among the general aspects of ethical AI for all, some International and Multinational Guidelines, though not directly targeting AI originally, are tackling various aspects that still apply to it. Industry initiatives, such as the IEEE Global Initiative on Ethics of Autonomous and Intelligent Systems (Shahriari & Shahriari, 2017) has produced comprehensive guidelines for ethically aligned design, focusing on how AI systems can respect and protect human rights. International efforts, like OECD Principles on Artificial Intelligence (2019) was adopted by over 40 countries, and outline standards for responsible and trustworthy AI, emphasizing respect for human rights, transparency, accountability, explainability, education and digital literacy, fairness, robustness, and security. Building on these, the G20 AI Principles were endorsed by leaders of the G20 countries, and further emphasized among other things, the notion of protecting human well-being. The AI Act, approved by the European Parliament on March 23rd, 2024 (AI Act, 2024)establishes a comprehensive regulatory framework for overseeing the development, deployment, and application of artificial intelligence across European countries. In its framework, it specifically addresses children and demonstrates the EU's proactive approach to regulating and safeguarding AI for children.

However, those do not specifically tackle the nuances of children's cognitive development. It is encouraging to observe significant organizations with regulatory influence turning their attention towards children's specific needs. Unfortunately, as revealed in our discussions, the principle of Safeguarding Cognitive, Social, and Emotional Development — vital for safeguarding children's holistic development against AI's potential negative impacts — has received less focus than others. The United Nations Convention on the Rights of the Child (United Nations Convention on the Rights of the Child, 1989), while not AI-specific, is a pivotal international treaty that champions children's protection and developmental rights, offering a foundation for AI regulations concerning childhood. Furthermore, the Children's Online Privacy Protection Act (Children's Online Privacy Protection Act (COPPA), 1998) by the US Congress though not directly targeting AI, still proves relevant as it governs the collection of personal information from children under 13 by digital services, which is an essential aspect of safety in regulating AI's development for children. Meanwhile, the General Data Protection Regulation (General Data Protection Regulation (GDPR) – Legal Text, 2018) was adopted in 2018 by the European Parliament and applies to all individuals. The Age-Appropriate Design Code (Age-Appropriate Design Code, 2020), enforced by the UK's Information Commissioner's Office, delineates 15 standards for age-appropriate digital services accessible to children. Although its jurisdiction is the UK, it inspires regulations globally and serves as a benchmark for designing child-safe digital services. Additionally, it underscores the importance of tailoring approaches to cater to the varied developmental stages of children.

Published in 2021, UNICEF's Policy Guidance on AI for Children underscores the principles of inclusion, fairness, privacy, safety, and the empowerment of children in the digital age, detailing nine key requirements for child-centric AI (UNICEF, 2021). UNICEF has outlined nine principles for child-centered AI in their Policy Guidance on AI for Children: 1) Support children's development and well-being, 2) Ensure inclusion of and for children, 3) Prioritize fairness and non-discrimination for children, 4) Protect children's data and privacy, 5) Ensure safety for children, 6) Provide transparency, explainability and accountability for children, 7) Empower governments and businesses with knowledge of AI and children's rights, 8) Prepare children for present and future developments in AI, 9) Create an enabling environment. Though it's encouraging that "Support children's development and well-being," is one of the nine principles, it is crucial to build on this and create more concrete guidelines for product developers as well as policymakers. Similarly, Common Sense Media's AI Initiative incorporates these considerations under the Kid's Safety Principle, focusing on protecting well-being.

Recognizing the intricate relationship between the environment and skill development, first we must contemplate the potential implications of these changes for child development. Specifically, we must ask ourselves: What will these transformations mean for the development of children? In the AI era, safeguarding children's development necessitates proactively identifying potential risks by leveraging insights into child development and preempting the possible negative impacts of technology on their lives.

## 1.5. Our Approach

In response to the urgency to anticipate the impact of AI-based technology on child development for proactive regulations, we consulted with experts from relevant domains. We reached out to specialists in AI, technological product development, as well as specialists in child development, conducting interviews with 16 renowned experts. They volunteered their time and shared their insights on future AI trends in AI, as well as how this might influence children both short and long term (See Annex 1).

The interviews averaged 105.4 minutes each, totaling 14 hours and 3 minutes. Those experts are cited as contributors in this paper, and were all contacted for validation before publication, 10 volunteered to review the report before its publications (See Annex 1).

# 2. ANTICIPATED AI USE FOR CHILDREN

Experts in AI have collectively anticipated the rapid development, deployment, and integration of AI in children's lives, leading to its pervasive nature in their environment. With recent leaps in generative AI progress, it will soon be either integrated and improved in existing solutions or incorporated into new ones, along three main domains that we will address here:

1) Entertainment
2) Education
3) Conversational agents

AI initially penetrated the **entertainment sector** through streaming, video games, and social media, making it a logical and reliable starting point for anticipating AI's impact on children and adolescents. Additionally, **education**, primarily through EdTech applications, has progressively integrated more AI tools, and the advent of generative AI promises to further transform the educational landscape. Generative AI is also revolutionizing the field of **conversational AI agents**—programs that can engage in conversation with humans such as robot, ChatGPT or digital voice assistant—making them more effective and prominent. This list is not exhaustive but reflects some of the most imminent and direct changes to children's environments, although AI has also infiltrated other domains, where it is expanded at an increased rate, such as health, transportation, and communication.

For each domain, we first discuss how AI will be improved or integrated into their applications. Then, we discuss what those changes might entail for children's development, underscoring both their potential and inherent risks. Children's perceptions, experiences, and interactions with their surroundings are constrained by the immaturity of their cerebral development and still limited life experiences. These components collectively influence how children engage with and make sense of the world around them, offering insights into their cognitive and socio-emotional development. Furthermore, this highlights the critical need to acknowledge children's perspectives, which often differ greatly from adults', when designing experiences for them, ensuring that these align with their current capacities.

## 2.1. AI in Entertainment, Leisure and Social Media: Enhancing Experiences and Addressing Challenges for Children

AI has deeply penetrated the entertainment sector over the past few decades and offers a wider view on the interaction of youth with technology. Children often engage with AI through sophisticated algorithms that **personalize content** in **streaming services** and **social media** by curating content of interest to increase engagement. AI also enhances experiences in **videogames** through dynamic difficulty adjustments. These interactions with content are bound to become more immersive with the development and deployment of **Immersive Virtual Reality (IVR),** which is already used in entertainment to transport players to alternate realities, like videogames, and in education to create more interactive learning experiences. These advancements underline the transformative potential of AI in entertainment, enriching user experiences through personalization, interactivity, and immersive storytelling.

The field of research has extensively examined the impact of digital media on children's development. Building on these findings, we reflect upon how AI has the potential to enhance cognitive and socioemotional development by building on existing solutions, continuously improving them, and by creating new ones. Then, we discuss the developmental considerations crucial to guide the further integration of AI, taking into consideration children's protracted development.

## 2.1.1. AI's potential for child development

**Enhancing cognitive development with AI**

The entertainment sector, with its increasingly personalized content and interactive elements, has the potential to be more stimulating and provide a richer learning environment. In terms of language acquisition in young children, exposure to television programs alongside parental involvement in co-viewing and interactive questioning have been shown to enhance learning, underscoring the ongoing importance of active engagement in language development (Strouse et al., 2018). Additionally, "fake interactions," such as engagement pauses in programs like Dora the Explorer, have been found to boost vocabulary learning in preschool children (Linebarger & Vaala, 2010). Generative AI could significantly enhance these benefits for **vocabulary expansion**, as evidenced by studies demonstrating increased learning and retention with interactive AI in TV shows (Y. Xu et al., 2022; Y. Xu & Warschauer, 2020). Regarding sensory development, smart toys employing augmented and mixed reality could offer **enhanced sensory and motor experiences compared to traditional tablets**, addressing the current limitations of engaging in digital worlds (Tang & Tewell, 2015). For instance, toys like soccer balls linked to apps or Legos augmented with AR could merge interactive digital media elements with physical experiences. This integration helps address concerns about the diminishing physical activity due to the allure of digital screens. Moreover, such toys could promote physical activity through exergames—video games that combine exercise and gameplay—which have been shown to aid in reducing BMI (Staiano et al., 2018).

Focusing on cognitive development, video games can have a significant impact, as evidenced by a report by the Entertainment Software Association (ESA, 2024) which notes that the U.S. alone has 227 million gamers across all age groups. These gamers engage across different platforms, including smartphones, game consoles, and computers, ranked in order of popularity. Engaging in some types of video games has been shown to **enhance top-down attention and spatial cognition**, offering promising insights for perceptual enhancements (Bediou et al., 2018; Choi et al., 2020). Nonetheless, the impact of gaming is nuanced due to the diversity of games which now include a wide array of genres and gameplay mechanics (Dale, 2020).

The evolution of the video game industry is closely linked with advancements in AI. Of note, AI allows the creation of more sophisticated and dynamic game environments that ensure that games remain both engaging and challenging—which aligns with the concept of Flow—for players by dynamically modifying the game's difficulty based on the player's performance (Bakkes et al., 2009; Sepulveda et al., 2019; Skinner & Walmsley, 2019; Torrado et al., 2018). Studies in education have found that when there is an optimal balance between a student's skill level and the challenge presented, students are more likely to enter a flow state. This immersive state not only enhances engagement but also significantly increases the likelihood of improved learning outcomes (Hamari et al., 2016; Mandhana & Caruso, 2023). Furthermore, if the AI can predict and generate guidance in a way that properly scaffolds the learning experience without providing excessive assistance, it would align with the educational concept of the Zone of Proximal Development. This facilitates optimal learning by matching challenges and scaffolding learning to suit the player's evolving capabilities (Bakkes et al., 2009; Sepulveda et al., 2019; Skinner & Walmsley, 2019; Torrado et al., 2018). With AI advancements, games are increasingly capable of assessing and adapting to a player's skill level, presenting significant learning opportunities (Bork, 2012). Due to their inherently motivating nature, video games have successfully been leveraged for therapeutic interventions, potentially reducing symptoms in children with Attention Deficit and Hyperactivity Disorder (ADHD) or with Autism Spectrum Disorder (ASD) (Jiménez-Muñoz et al., 2022; Peñuelas-Calvo et al., 2022).

**AI as a tool for socio-emotional development**

A recent study by Bain & Company involving 25 game industry executives anticipates that generative AI will revolutionize the industry by enhancing the complexity and interactions of NPCs (non-player characters). An NPC is a character in a video game that is not controlled by a player but is instead operated by the game's software. Given that **prosocial video games** have been associated with increased prosocial behaviors and reduced violence (Greitemeyer, 2022; Li & Zhang, 2023), and **multiplayer** online games enhance soft skills (Pagel et al., 2021), the sophisticated interactions with NPCs could offer new avenues for social skill development. Additionally, the use of Immersive Virtual Reality (IVR) has been shown to positively influence attitudes towards individuals from outgroups, such as individuals from other gender or ethnics (Fox et al., 2013; Peck et al., 2013; Yee & Bailenson, 2007). It is possible then to assume it could serve as a tool to teach about empathy or bias awareness.

**Social media**, often scrutinized for potential risks to adolescents, also play a crucial role in their social development. They can facilitate the establishment and maintenance of friendships by, for example, offering opportunities **for emotional and social support** and shared activities (Angelini et al., 2022; Bukowski et al., 2009). Social media can also help adolescents stay in touch with their support networks, which is especially important for kids from marginalized groups, such as LGBTQ+ adolescents who may be reluctant or unable to discuss their identity with caregivers, and where online support can represent a lifeline (Craig et al., 2021; Lucero, 2017; Selkie et al., 2020). AI also holds the potential to address current limitations in social media, enhancing content moderation and potentially reducing exposure to harmful content and interactions, though researchers in this field highlight the technical and ethical complexities that this raises (Hakimi et al., 2024).

Finally, in the **medical field,** a systematic review of 17 studies focusing on medical procedures suggests that immersing patients in virtual reality can significantly **reduce pain and anxiety for children and adolescents during and after medical interventions** (Eijlers et al., 2019). IVR shows promise in preparing children for upcoming medical procedures by immersing them through their different steps (Gold et al., 2021; Stunden et al., 2021). The effectiveness of IVR extends to treating phobias and social anxiety in adults through Virtual Reality Exposure Therapy (VRET), with systematic reviews citing encouraging outcomes (Botella et al., 2017, 2017, 2017; Horigome et al., 2020). Recently, this approach has been explored for treating phobias in children. Although research in this area, particularly randomized controlled trials, remains scarce (Kothgassner & Felnhofer, 2021), they remain nonetheless an interesting avenue of treatment.

### 2.1.2. Developmental Considerations

Similarly to the potential benefits AI-integration to products can offer, child developmental considerations and responsible decision-making should guide the development of future applications. By building on existing research and data in children and digital media, learning from past mistakes in AI, and embracing speculative future thinking, we can better anticipate and avoid future pitfalls.

**The trade-offs of prolonged screen times**

Children and adolescents are interacting with digital media at increasingly younger ages, which represents more than four hours a day to such activities before reaching the age of 14 (Radesky & Christakis, 2016; Rideout, 2021). This shift away from other leisure activities is already causing harm to their vision and overall well-being, through physiological aspects such as sleep and sedentariness. Increasing the number of products, applications, and activities infused with AI has the potential to lead to even higher levels of consumption, as suggested by the current trends (Rideout, 2021), further diminishing engagement in other crucial, varied, and rich experiences, such as outside free play with other children.

High digital screen use and a low level of outdoor activity among children are linked to **myopia**, or nearsightedness (Jones-Jordan et al., 2012). In Asia, the situation is particularly alarming, with up to 73% of children aged 12 to 18 affected in South Korea (Grzybowski et al., 2020). Screens contribute to myopia as they require continuous close-up focusing and exposure to blue light. This exposure is more damaging to children due to the physiological immaturity of their eyes, which are less capable of filtering out harmful light, increasing their vulnerability to visual stress and structural changes (Artigas et al., 2012). Looking forward, experts estimate that by 2050, half of the global population will suffer from myopia, with 10% facing severe implications (Holden et al., 2016). The long-term consequences of myopia extend beyond poor vision, as it also increases the risk of serious eye conditions in adulthood, such as maculopathy, retinal detachment, glaucoma, early cataract formation, and potential blindness. These conditions not only impact individual health, neurological development and quality of life but also impose significant economic burdens due to increased healthcare needs and lost productivity.

As for the physiological impact of digital media consumption, young people tend to play video games or scroll through social media during the early hours of the night, **sacrificing sleep time and quality** (Hale & Guan, 2015; Pirdehghan et al., 2021; Scott & Woods, 2018). Sleep is an essential phase for brain development and functioning, providing critical time for the consolidation of memories, processing of experiences, and restoration of neural pathways (Potkin & Bunney, 2012). This chronic lack of sleep exposes children and adolescents to negative cognitive consequences, such as issues with attention, memory, and learning (Peracchia & Curcio, 2018; Weaver et al., 2010; Wolfe et al., 2014), as well as affective issues, such as anxiety and depressive moods (Bastos et al., 2023; Pires et al., 2012; Talbot et al., 2010).

More appealing AI-infused entertainment could also further exacerbate an already excessive **sedentary lifestyle**, which is in turn associated with a higher risk of **physical health issues** (Costigan et al., 2013). Screen media use is a significant factor in the global child and adolescent **obesity** epidemic (Robinson et al., 2017), with positive correlations between screen time and weight in children (Hesketh et al., 2007; Robinson et al., 2017). Additionally, poor physical activity is linked to **higher levels of anxiety and depressive symptoms** (Brown et al., 2013; McMahon et al., 2017). Conversely, a review of existing interventions that increase physical activity has been shown to reduce anxiety in healthy patients (Rebar et al., 2015). Some evidence also suggests that physical activity has positive effects on cognition and academic performance in children (Donnelly et al., 2016).

Another way spending more time-consuming digital media might negatively impact children's development is by reducing exposure to foundational experiences at specific ages during critical periods of development. In babies and toddlers, this is even more crucial since they seldom learn from digital media exposure, a phenomenon known as the **video-deficit effect** (Anderson & Pempek, 2005; Krcmar, 2010; Linebarger & Vaala, 2010). This concept is crucial when developing solutions for **children under 3 years old**, where increased interactivity through touch screens does not always translate into learning (Sheehan & Uttal, 2016). This is partly due to the high reliance of children under 3 on social contingency, meaning they need the responsiveness of a caregiver to their actions for optimal learning (Dunst et al., 2008; Tarabulsy et al., 1996). An additional key concept is that children develop dual representation gradually between the ages of 3 and 5 years old. **Dual** **representation** posits that for children to fully understand and learn from symbols, for instance, a flashcard of a dog, they must simultaneously hold two mental representations: the symbol as a physical object in its own right (the flashcard) and the symbol (the dog) as a representation of something beyond itself (DeLoache, 1989, 2000). When interacting with screens, they often put more focus on the screen itself than the symbol it represents (Anderson & Pempek, 2005; Krcmar, 2010; Linebarger & Vaala, 2010). A longitudinal cohort study demonstrated a persistent link between early screen media exposure and cognitive development (Pagani et al., 2010). Specifically, the study found that each additional hour of TV exposure at the age of two was associated with a 7% decrease in classroom engagement and a 6% reduction in math proficiency by the fourth grade. Therefore, a potential risk of AI solutions in entertainment could create **overreliance on unproven methods that replace crucial language development activities before age 3** (Putnick et al., 2023). While GenAI in educational shows hold promises for vocabulary development, the **importance of rich social interaction** for broader communication and social skills development remains paramount. With older children GenAI has the potential to make digital media more engaging by creating interactions that encourage children to be active learners, rather than passive consumers (Y. Xu et al., 2022; Y. Xu & Warschauer, 2020). However, for AI to hold that promise, the emphasis has to be on consciously creating enriched experiences that support learning, rather than overly easy digestible content that merely increases engagement time.

In both babies and preschool-aged children, increased time spent engaging with digital media might lead to **reduced engagement with physical objects and rich sensory experiences**, which are critical at this age. Current digital experiences often lack the tactile and motor engagement provided by physical manipulation (Crescenzi et al., 2014). Longitudinal studies have reported associations between prolonged screen time and **decreased sensory integration and fine motor skills** (Cadoret et al., 2018; Heffler et al., 2024; Suggate & Martzog, 2021), suggesting that this shift away from rich sensory experiences could have implications for healthy sensory-motor development (Lin et al., 2017). Indeed, sensory integration is associated with higher-order cognitive functions related to cognitive control, such as attention, executive functioning, and emotional regulation (Chaddock et al., 2011; Chang et al., 2022; Hilton et al., 2007; Kojovic et al., 2019; Masten et al., 2012; Pangelinan et al., 2011; Schmahmann & Pandya, 2008).

**Children still developing minds**

The trade-off of spending more time with digital media represents a risk of lacking some of the variability and depth of experiences. Further, the experiences provided by those applications are not always adapted to children. Children and adolescents' cerebral immaturity or still-developing mental representations of the world might make them specifically vulnerable to some AI uses. Here, we specifically address children's progressive understanding of privacy and the specific risks of being exposed to harmful content. We highlight the potential impact of attention-capturing algorithms on delayed gratification and on the reward system, as well as their influence on attention and executive functioning. Finally, we explore the potential adverse effects of IVR on children.

The notion of **privacy** is a perfect illustration of the difficulty for young people to make informed decisions considering the complexity of the implications and abstraction skills they require. It represents a common concern in the entertainment and AI sectors for both experts and parents, especially regarding social media and Immersive Virtual Reality (IVR) (Kelly, G., Graham, J., Bronfman, J., & Garton, S., 2022). It is crucial to take into account that **children's understanding of privacy** is not the same as adults', and their development of this concept is progressive with age, and also varies depending on their personal circumstances, such as socioeconomic status (SES) (Livingstone, S. Stoilova, M. and Nandagiri, R., 2019). Children under 7 lack the abstract understanding of concepts like 'privacy' and 'safety' (Chaudron et al., 2018). While by age 11 their comprehension improves, they still lack judgment in applying these concepts to practical situations (Kumar et al., 2017). For instance, children under 11 are more likely to share their data on websites that contain warnings for age-inappropriate content as this elicits their curiosity (Miyazaki et al., 2009). Even during adolescence, their decisions often prioritize immediate gratification over the consideration of uncertain risks in the future, despite a more mature understanding of those concepts (Youn, 2009; Yu et al., 2015). While digital literacy is important, they are still immature and their difficulty to exercise their judgment when making decisions for long-term effects show that it is not sufficient in protecting them. Furthermore, they lack a safe space to practice these concepts of privacy without facing repercussions. The role of adults and policymakers in protecting their privacy in the context of this immaturity is essential (Livingstone, S. Stoilova, M. and Nandagiri, R., 2019).

Another potential risk involves the future capability for users to self-generate content, potentially complicating the control of age-appropriate viewership for children and adolescents. Exposure to content that is not age appropriate is associated with behavioral and emotional changes, such as heightened aggression (Calvert et al., 2017; Greitemeyer, 2022) or the adoption of inappropriate and unsafe sexual behaviors (Browne & Hamilton-Giachritsis, 2005; Massey et al., 2021).

Due to their still-developing brains, young children may struggle to comprehend the implications of their actions, making them more likely to be attracted to **harmful content** in the digital world. Additionally, because of this immaturity, they are especially vulnerable to the negative impact of such harmful experiences on their development. Content curation through AI represents such a potential risk. The first is that children's ability to regulate their emotions develops progressively and only reaches full maturity in early adulthood (Casey et al., 2019). They rely more on immediate gratification and might, in turn, have more difficulty regulating their use of digital tools, especially as AI's ability to curate highly rewarding content improves. Although individual differences in impulsivity mediate the ability to delay rewards (Minear et al., 2013; Sanbonmatsu et al., 2013; Shih & Chuang, 2013), this might be particularly true for adolescents. As previously mentioned, according to the dual systems model, adolescents are biased to respond vigorously to rewarding and novel experiences due to the more rapid maturation of the reward system relative to the cognitive control system (Casey et al., 2008; Ernst et al., 2006; Luciana & Collins, 2012; Steinberg et al., 2018). Therefore, adolescents may not be equipped to make proper decisions effectively in a reward-driven context due to their developmental cerebral constraints (Luciana & Collins, 2012).

Additionally, higher use of mobile technology could generate higher need for **instant gratification** (Wilmer et al., 2017). Individuals who are heavier users of mobile technology are more likely to accept smaller, more immediate rewards than to wait for a more substantial but delayed one (Wilmer & Chein, 2016). Researchers have found that after a 3-month exposure to smartphones, non-users became more immediacy-oriented in a delay discounting measure suggesting that heavy smartphone usage can causally **reduce an individual's capacity** (or at least tendency) **to delay gratification** (Hadar et al., 2015).

Behavioral addictions are linked to impulsivity, and they might be particularly linked an inability to delay gratification (Li et al., 2016; Munno et al., 2016). These addictions involve repeated dysfunctional behaviors that do not require the ingestion of addictive substances (Goodman, 1990; Griffiths, 1996), such as Online Gaming Disorder, which is included in the DSM-5 (American Psychiatric Association, 2013). Two meta-analyses focused on neural changes in individuals with Internet Gaming Disorder have reported abnormalities in frontostriatal and fronto-cingulate circuits, as well as dysfunctions in the prefrontal lobe in individuals with this disorder (Meng et al., 2015; Yao et al., 2017). These structures are also related to emotion dysregulation in other addictive disorders and contribute to the compulsive use of screen devices in general (Feil et al., 2010; Lüscher et al., 2020).

High social media use has been linked to diminished attentional control in adolescents by reducing their ability to maintain focus and resist distractions (Siebers et al., 2022). This issue arises as notifications and media use may divert their attention from their immediate tasks to previously encountered online content or interactions (Stothart et al., 2015). Receiving notifications during a math task, for example, can significantly affect both accuracy and reaction time, with more pronounced effects in adolescents (15 years old) compared to adults (Whiting & Murdock, 2021), suggesting a higher vulnerability for this group. Although batching notifications can lead to reduced stress levels by minimizing constant distractions, for those with a high fear of missing out, it may have the contrary effect, intensifying the urge to compulsively check their devices (Fitz et al., 2019; Liao & Sundar, 2022; Rozgonjuk et al., 2019). Furthermore, screen media also impacts attention skills by reducing the amount of sleep adolescents get every night (De Oliveira et al., 2020). An intervention involving over 500 adolescents aged 12 to 19 years showed that discontinuing digital media use after 9 p.m. led to earlier sleep onset and longer sleep duration (Perrault et al., 2019). This resulted in enhanced daytime vigilance, with personal biological factors playing a mediating role (Perrault et al., 2024).

Executive Functioning (EF) is a set of cognitive skills including inhibitory control (ability to suppress competing goal-irrelevant information), working memory (ability to actively maintain or update goal relevant information), and flexibility (ability to quickly adapt to changing circumstances) (Miyake et al., 2000). Executive functioning develops all the way through adolescence to early adulthood. During adolescence, heavy digital media usage has been linked to **diminished executive functioning (EF)**, particularly through increased media multitasking among adolescents (Alho et al., 2022). This behavior is associated with lower scores on standardized tests measuring academic performance in English and math, indicating a correlation between **multitasking** and reduced academic success (Baumgartner et al., 2017; Liu et al., 2020; Van Der Schuur et al., 2015). The inclination towards digital media multitasking and its resultant impact on attention and executive functioning seems to be influenced by individual cognitive profiles. Specifically, adolescents diagnosed with Attention Deficit Hyperactivity Disorder (ADHD) are more prone to engage in multitasking with digital media, which can further compound attentional difficulties (Baumgartner et al., 2017). Additionally, participants demonstrated impaired inhibitory capacities following smartphone use, as opposed to watching a documentary for a similar duration (Jacquet et al., 2023).

Though humans have always engaged with **fantasy** through various forms of expression and mediums that have evolved over millennia, from cave painting to folklore and performing arts, IVR marks an evocative change due to its immersive qualities. IVR is processed by the brain closer to an actual experience than a media one, even in adults (Bohil et al., 2011). While Immersive Virtual Reality (IVR) use by adults and the appropriateness of what should be experienced in the metaverse is already an important question, it is even more so for children. IVR promises to become increasingly more widespread in the upcoming years, with proliferation of constructors and products entering the market, including the VisionPro from Apple earlier this year (Z. Zhang et al., 2023) (REF). This should raise questions about how this will impact children's cognitive development and overall well-being. For a thorough investigation of the topic, read Common Sense Media report (Common Sense Media, 2018).

Children first develop the ability to **distinguish between reality and fantasy** through their assessments of what is real and what is not, often categorizing events they have not personally experienced as improbable (Woolley & Ghossainy, 2013). By the age of 5, they can discern the difference between reality and fantasy portrayed on television (Mares & Bonus, 2019; Woolley & Ghossainy, 2013; Wright et al., 1994). Entering middle childhood (around ages 7–8), children enhance their evaluative skills, determining the reality of a situation based on its feasibility or likelihood in the real world. For instance, IVR has been shown to potentially influence attitudes, behaviors, and physiology in adults (Woolley & Ghossainy, 2013). Furthermore, children tend to perceive IVR experiences as even more realistic compared to adults (Woolley & Ghossainy, 2013).

Drawing from this, it is worth asking how a child will react to experiencing the virtual appearance of a dinosaur, bear, or ghost in their living room, and how that might potentially affect the development of their concept of reality-fantasy. Reality-fantasy confusion in children is associated with heightened nightmare fears, particularly in younger children (Zisenwine et al., 2013). Indeed, three in four children aged 4 to 12 experience nightmares (Bauer, 1976; Gordon et al., 2007; Muris et al., 2001), with most children attributing their fears to exposure to negative information (Muris et al., 2001). In a large-scale study, children attributed a large portion of both their pleasant dreams and their nightmares to their digital media consumption (TV and video games) (Van Den Bulck, 2004), underscoring its influence. In this context, the immersivity of IVR might increase nightmares and anxiety if exposed to undesired content, even one that might seem ordinary to adults, as what constitutes scary content is often misinterpreted and can be unnoticed by adults. It is also worth considering how IVR might influence children's understanding of the distinction between fantasy and reality.

Furthermore, the immersive nature of experiences in VR increases psychological presence. In young adults, playing a video game in VR rather than traditional desktop format leads to playing more aggressively, and increased self-reported aggressive feelings and elevated heart rates (Persky & Blascovich, 2007, 2008). This **higher experienced psychological presence**, in turn, increased levels of anger after play (Lull & Bushman, 2016). In this context, the **exposure to violence in VR** seems to have a different and **more pervasive impact** on the player than in traditional video games (Bailey & Bailenson, 2017).

Even positive experiences through IVR should raise serious questions considering children's cerebral developmental immaturity. Children can be influenced by seeing their **virtual doppelgangers**, avatars that closely resemble them, engaging in experiences through IVR. For instance, elementary aged children created **false memories** after watching their avatar swimming with orca whales, believing it truly happened to them (Segovia & Bailenson, 2009). This potential confusion with reality is also reported in adults (Bonnail et al., 2024). This was also the case when watching videos on TV for children younger, though no longer the case in elementary children, reinforcing the idea that impacts of content experienced through IVR should not be interpreted through our current knowledge of how children interpret other screens.

Regarding the influence of avatars, this area warrants further examination, particularly as research among young adults begins to reveal significant potential impacts. The attitudes of users are influenced in various ways by the appearance of their avatars. When the avatar closely resembles the user, there is a noticeable increase in skin conductance, indicating a stronger preference and **identification with the avatar** (Blascovich & Bailenson, 2011; Fox et al., 2009, 2012). For instance, users tend to favor brands they have seen on their virtual doppelgänger over those presented on different avatars (Ahn & Bailenson, 2011). Moreover, the concept known as the Proteus effect highlights **how an avatar's appearance can influence a user's attitudes and behaviors in the real world** (Yee & Bailenson, 2007). Studies suggest that embodying a tall avatar was shown to boost confidence, while engaging as a hypersexualized female avatar induced feelings of self-objectification, and experiencing an avatar of a different race can reduce implicit biases (Fox et al., 2009; Peck et al., 2013; Yee & Bailenson, 2007). It remains a complex question whether this could foster development of empathy or complicate the development of self-identity and personal sense of self. Children are progressively discovering and learning about who they are from the experiences they live, the way they react to them, and share them with others. In this context it remains an essential question of how metaverse experiences will influence their sense of self and personal memories within such contexts.

Finally, the integration of GenAI in Non-Playable Characters (NPCs) introduces **complex dynamics in player-NPC relationships**. Historically, research on NPCs has primarily concentrated on their design, aiming to make them more realistic and engaging (Daviault, 2012; Mallon & Lynch, 2014). However, a significant research gap exists regarding the study of **players' emotional attachments to NPCs**. Initial studies suggest that players can form deep, meaningful emotions towards NPCs, including feelings of protectiveness and even love (Bopp et al., 2019; Burgess & Jones, 2020; Headleand et al., 2016). The intensive effort to enhance Non-Playable Characters (NPCs) and boost player attachment, juxtaposed with the low number of investigations into the emotional effects on players, might highlight a misplaced set of priorities. With the anticipated use of GenAI in NPCs, the potential for even deeper bonds and stronger emotional connections is highlighted, akin to interactions with AI bots (Boine, 2023). Such advancements could lead to significant emotional distress for players, especially when NPCs face harm or death within game narratives. Furthermore, relationships with NPC are likely to parallel challenges observed with AI bots, including **dependency** on NPCs, receiving potentially harmful advice, or **negative impacts on real-life relationships** (Boine, 2023). This underscores the need for comprehensive research to understand and mitigate the potential emotional and psychological effects of deepened player-NPC relationships facilitated by GenAI technology. Priority should be placed on ensuring the emotional safety of players over the creation of more immersive and engaging experiences.

**Interpersonal mediating factors: How AI impacts different people differently**

The interaction between AI-solutions and children or adolescents will have different implications that will be mediated by interpersonal factors that can be either protective or make some populations even more vulnerable. The impact of some applications appears to be different according to gender, age, race, mental health and neurodevelopmental profiles, as well as life experiences.

When it comes to the impact of aggressive content on aggressive behaviors, a broad field of study indicates that **violent video games may increase hostile attributions, negative affect, and physiological arousal.** And this relationship appears to be moderated by personal factors such as age — with a notable U-shaped correlation peaking at age 14 (Burkhardt & Lenhard, 2022), gender — with **boys** exhibiting more aggressive cognitions and behaviors than girls — and personality traits (López-Fernández et al., 2021).

On the other hand, **girls' body image, self-esteem and mental health** appear to be disproportionately affected compared to boys. Social media may specifically contribute to increased body image concerns by putting the focus on physical appearance through exposure to idealized images and quantifiable indicators of approval, as well as their own appearance (Choukas-Bradley et al., 2022; Scully et al., 2023). Social media's impact on mental health in teens is also influenced by age, self-esteem, and the nature of the user's engagement with social media — whether active or passive (Blomfield Neira & Barber, 2014; Frison & Eggermont, 2016; Thorisdottir et al., 2019; Tsitsika et al., 2014; P. Wang et al., 2020). Of note, the representation of female characters in video games often reflects significant gender disparities, which can have detrimental effects on the well-being of female gamers. Studies indicate that female characters are frequently portrayed in a subordinate role to male protagonists, objectified, and hypersexualized with unrealistic body proportions, which both reinforces sexist attitudes among male gamers, and negatively impacts female self-esteem and body image during crucial formative years (Cooke et al., 2012; Gestos et al., 2018; Kaye & Pennington, 2016; Paaßen et al., 2017).

Video games have become a significant influence in societal culture, unfortunately perpetuating negative racial biases based on stereotypes held by white individuals about people of color. White males are overrepresented (Williams et al., 2009), while African Americans are frequently depicted in derogatory and unflattering ways that reflect these prejudiced views (Burgess et al., 2011; Dickerman et al., 2008). This portrayal can negatively affect players' race-related perceptions, particularly when they embody characters that are racially stereotyped in these negative ways (Behm-Morawitz et al., 2016). Similar issues of representation have also been reported in AI image generators, which often produce racist and sexist results.

Mental health also influences the use of such AI-driven applications, which then can lead to differential impacts on more vulnerable populations. Longitudinal research has indicated that **depressive symptoms can often precede problematic use of social media** (Puukko et al., 2020; Raudsepp & Kais, 2019; Seabrook et al., 2016), suggesting that depression may lead to increased social media use as a form of compensation, which in turn may worsen depressive symptoms through unhealthy online behaviors (Appel et al., 2016; Raudsepp, 2019; Seabrook et al., 2016). Further, limiting social media usage can result in participants reporting lower levels of loneliness and depressive symptoms (Hunt et al., 2018) and reduce stress levels, particularly among those who previously engaged in excessive use of social media (Turel et al., 2018).

As previously mentioned, individuals who are neurodivergent might be specifically negatively impacted by AI algorithms which stimulate the reward system to increase engagement. Individuals with Attention Deficit and Hyperactivity Disorder (ADHD) are more susceptible to problematic gaming and **addictive behaviors**, with higher risks of developing gaming addiction (Bioulac et al., 2008; Masi et al., 2021). Video games can exacerbate inattentive symptoms in children with ADHD—a condition representing between 4% and 9% of the pediatric population (Salari et al., 2023). They also might be disproportionately represented among gamers, which could be attributed to the rewarding nature of games and the characteristics of ADHD itself (Scahill & Schwab-Stone, 2000).

## 2.2. AI in Education: Balancing Innovation with Careful Integration

Digitalization in education is not new, but AI promises to further revolutionize this sector (Cardona et al., 2023). The average number of Educational Technology (EdTech) tools used per district in the US went from 300 in 2017 to over 2,500 in 2023, with 65% of teachers incorporating digital tools into their daily interactions with students, as well as 70% of students utilizing them outside of school settings (Gallup Inc., 2019). This trend has likely intensified in the wake of the COVID-19 pandemic, further embedding digital tools in educational practices. Meanwhile, UNESCO released a report in 2023 urging countries to **better regulate and assess EdTech products that often lack proven efficacy** and reinforce the importance of social connection in learning. In this evolving context, advancements in AI bring new opportunities for improved tools that are more interactive, complex, and responsive, bridging some of the existing limitations (GEM Report UNESCO, 2023). However, experts in child development also caution that these advancements could increase the risk of replacing proven methods with new, less-effective ones.

### 2.2.1. AI's potential to revolutionize education

One of the obvious potentials of AI in education is its ability to reach more learners, regardless of their geographical location or economic background, offering the promise of a personal tutor with almost infinite knowledge at our fingertips. Enhanced online education could provide educational opportunities for more isolated communities and facilitate access to higher-quality content on a worldwide scale. Experts emphasized the potential for technology to improve **accessibility** by enhancing access to education and reducing disparities. By offering wide-scale curricula that are common across many students and schools, technology can facilitate access to diversified enrichment programs by cutting costs. This is especially significant in the United States, where higher socio-economic status (SES) school districts typically have more activities available and parents in these districts are often able to provide additional extracurricular activities for their children, leading to increased inequality in education (Braga et al., 2017; Kornrich, 2016).

**Personalized learning**

A growing trend in education is personalized learning, where AI tailors educational content to the individual's learning style and pace. These platforms use data analysis and machine learning algorithms to **adapt to the student's progress**, providing additional support or advanced challenges as needed. Thanks to GenAI, it could also tailor the content to students' and educators' interests to make it feel more relevant to them. The concept of using technology to tailor learning experiences dates back to the mid-20th century, with Skinner's development of the personalized learning machine in the 1950s (Skinner, 1968) His machine presented educational materials and provided feedback to learners at their own pace, laying the groundwork for today's adaptive learning technologies. The evolution from Skinner's early contributions to the current landscape highlights the enduring vision of personalized education through technology.

Indeed, AI can enhance student engagement through interactive educational solutions by dynamically adjusting difficulty levels and **gamified design**. Those learning tools often incorporate point systems, badges, and ranking, positing it will make learning more attractive and will boost motivation through rewards. These tools are frequently used to teach subjects like math, language, and science. Students generally report feeling more engaged, wanting to use these tools more often, and feeling more motivated when using such gamified learning tools (Oliveira et al., 2022; Smiderle et al., 2020). However, it is important to exercise caution regarding the over-gamification of education (see section on developmental perspective).

One of the most promising aspects of personalized learning lies in AI's ability to provide **immediate feedback**, which has the power to significantly catalyze cognitive and motor skill development. A 2020 meta-analysis of 435 studies highlighted the complexity of its impact, with several factors mediating its effectiveness in fostering learning (Wisniewski et al., 2020). There is consensus that corrective feedback significantly improves learning, while punishment, praise, and reward have low to medium positive effects (Hattie & Timperley, 2007). Specifically, formative feedback is highly effective for learning: it is non-evaluative, supportive, timely, and specific, responding to the learner's actions with accuracy verification, explanations, hints, or worked examples. Its effectiveness is mediated by individual learner characteristics and the type of tasks (Shute, 2008). Providing such formative feedback is complex and time-consuming, however, AI technology could help educators deliver more personalized and effective support, ultimately improving the learning experience. Including AI-driven feedback in the classroom has been shown to positively influence learning (Ajogbeje, 2023). For instance, robots providing immediate feedback to high school students improved motivation, engagement, and overall learning achievements (Al Hakim et al., 2022). The interactive nature of AI also has the potential to adopt a **Socratic approach to assessment**—by asking questions to encourage deep thinking and reflection, enabling students to demonstrate their understanding and reasoning skills—potentially resulting in increased learning outcomes (Sorvatzioti, 2012).

AI also holds tremendous potential in supporting **children with special needs** both within and beyond the classroom. Popular applications already use adaptive learning to personalize instructions, by increasing engagement through interactive design, providing individual feedback, and adapting to children with special needs in education. **Assistive technologies** can range from low-tech devices like adapted keyboards to AI-infused high-tech solutions, such as screen readers and assistive listening systems, and are essential for overcoming barriers to learning (Hersh & Johnson, 2008; Lynch et al., 2022). These technologies have been shown to potentially improve academic engagement, psychological well-being, and social participation (McNicholl et al., 2021). A study of secondary school students with disabilities in the United States found that assistive technologies are used to support deaf-blind students, students with visual impairments, students with learning disabilities, students with emotional/behavioral disorders, and students with speech and language impairments (Bouck & Long, 2021). Assistive devices have also been beneficial for students with intellectual disabilities and children with Down syndrome, aiding in the development of skills such as numeracy, speech, language, memory, and social interaction (F. H. Boot et al., 2018; Shahid et al., 2022). AI can also significantly enrich and facilitate learning by adapting content and enhancing learning strategies. Such applications can profoundly impact the lives of these children, making education more accessible and effective for neurodivergent students (Cunff et al., 2022).

**Academic engagement (AE)** is recognized as a crucial contributor to school achievement (Greenwood et al., 1994). Low engagement adversely affects a student's ability to learn, and simple monitoring interventions such as questionnaires and behavioral observations by teachers and students can enhance both engagement and learning (Schardt et al., 2019). With technological advancements, monitoring allows for more sophisticated tracking with the aim of enhancing it and providing valuable feedback to students and teachers to further improve student engagement (Carroll et al., 2020). This monitoring ranges from minimally invasive techniques, such as tracking student interaction with EdTech products in the U.S. (Hankerson et al., 2022), to more invasive methods involving physiological monitoring devices used in China (Wang, Yifan; Tai, Crystal, 2019).

More invasive **monitoring** measures involve using physiological signals such as heart rate, skin temperature, respiratory rate, oxygen saturation, blood pressure, and electrocardiogram (ECG) data, alongside eye-tracking systems (Bustos-López et al., 2022; Y. Wang et al., 2021). These technologies can be leveraged to detect shifts in engagement, attention and stress level, which can facilitate the adaptation of teaching platforms to better suit the learner's cognitive and emotional states (Apicella et al., 2022). An experiment conducted in the US in K-12 classrooms showed that the use of wearable mixed-reality smart glasses by teachers, which provided real-time analytics on students, demonstrated positive effects on student learning outcomes (Holstein et al., 2018). Of note, such technologies can be highly effective in identifying certain patterns, indicating its potential to play a pivotal role in the early identification of specific needs in children (Mengi & Malhotra, 2022). Early intervention, facilitated by such timely detection, is often crucial in improving the prognosis for children with special needs (Okoye et al., 2023). However, this also raises important ethical concerns regarding students' right to privacy, and striking a balance between the benefits and risks of personalized and adaptable learning content represents a significant challenge.

**Innovation in content creation and presentation**

Experts interviewed recognized the transformative potential of AI in **supporting teachers and educators**. Drawing on insights from the Amazon Web Services (AWS) 2023 report on AI's potential for enhancing efficiency in various industries (AWS, 2023), it bears similar potential in education primarily in aiding the creation, adaptation, and management of educational content and resources (Hassany et al., 2023). Teachers are starting to use AI in the classroom to enhance their efficiency and effectiveness. As of fall 2023, 18% of K-12 teachers in the US had reported using AI to adapt content to fit the level of their students and to generate materials. By the end of the 2023–2024 school year, 60% of US districts had planned to train teachers about AI use policy, primarily because they saw the potential for AI to make teachers' jobs easier (Diliberti et al., 2024).

AI-powered tools can also provide **innovative methods** and represent **a new medium** to learning. **Immersive Virtual Reality (IVR),** through Virtual Reality (VR) and Augmented Reality (AR) can transport students to different historical periods or biological environments, making complex concepts easier to conceptualize. A review of existing research into VR's potential in education indicates it as a promising avenue for presenting complex concepts, particularly for visually intensive content in higher education (Radianti et al., 2020). It allows for a multisensory experience and constitutes an additional medium, both of which are known to have a positive impact on learning (Mayer, 2002). The effectiveness of IVR as an educational tool will largely depend on its utilization by teachers. There exists a parallel between using VR to present material and the employment of slide show type presentations in classrooms. When utilized as a visual aid, slide presentations can render material more engaging, thereby benefiting student motivation and learning (Johnson & Christensen, 2011; Tangen et al., 2011). Its application in K-12 education also shows promise (Araiza-Alba et al., 2022). However, using VR with children represents specific challenges and considerations, as mentioned in the section on AI entertainment.

It appears that AI can positively be used as a tool to improve education and learning outcomes. However, for all the promises it holds, synergistic collaboration between educational scientists and AI developers will be crucial to ensure that AI-driven educational tools are effectively tailored.

## 2.2.2. Considerations to ensure effective AI integration in education

**EdTech's existing limitations**

The field of EdTech remains largely unregulated, allowing products to enter the market without mandatory evaluation to prove their efficacy. To address this gap, researchers in the US developed the EdTech Evidence Evaluation

Routine (EVER) based on the four pillars of learning: active engagement (encouraging participation), meaningful learning (connecting new information to real-world contexts and prior knowledge), social interaction (promoting collaboration), and iterative learning (allowing for repeated practice and feedback) (Hirsh-Pasek et al., 2015). Strikingly, an analysis of the top-rated educational apps on the Apple Store revealed that most provide minimal learning value (Meyer et al., 2021), and more than half of the apps assessed using EVER exhibited low-quality design (Kucirkova et al., 2023). Despite the availability of systems such as the EVER system, its use remains limited, contributing to the **persistence of low-quality educational apps in the market**.

The limited efficacy of some EdTech solutions can be attributed to their design limitations, specifically the risk of **prioritizing engagement over educational value**, which can lead to superficial learning. This issue is common in EdTech, where the primary metric often revolves around measuring child engagement based on the time spent using an app, rather than the amount of knowledge acquired (Kucirkova et al., 2023). To increase engagement time, gamification is frequently used, as it boosts motivation to use the product.

In this context, it is crucial to distinguish between intrinsic motivation (engaging in an activity because it is inherently enjoyable) and extrinsic motivation (doing something for a separable outcome), according to Self-Determination Theory (SDT) (Ryan & Deci, 2000). The issue faced by gamification is that it creates intrinsic motivation to practice the exercise by making it more appealing, which then moves the motivation towards being extrinsic, since it is no longer about the learning. This means students engage for the rewards, further **displacing intrinsic motivation away from learning**. This shift can lead to negative learning outcomes, as illustrated by research showing that students in a gamified course exhibited less motivation, satisfaction and empowerment over time, and ultimately obtained lower final exam scores compared to a non-gamified class, highlighting the potential drawbacks of gamified approaches (Hanus & Fox, 2015). SDT's sub-theory, Organismic Integration Theory, details how extrinsic motivation can be internalized and integrated into an individual's goals, making it feel more self-determined (Ryan & Deci, 2000; Vansteenkiste & Deci, 2003). Therefore, designing for internalized extrinsic motivation may be most effective for sustained engagement and learning (Cerasoli et al., 2014). It also parallels the need for feedback that furthers learning outcomes, such as formative feedback, rather than gamified ones such as points or trophies.

Additionally, though personalized learning can lead to improved learning, up to this point, scientists underscore the **insufficient profound understanding of the underlying pedagogies and the learning process** (Bartolomé et al., 2018). Bridging the gap between academic research and practical educational applications is essential in addressing these challenges (Bernacki et al., 2021).

## Questioning Potential Unintended Ripple Effects of Digital Learning on Children

Development and learning are complex, non-linear processes that emerge from the continuous interaction of multiple, interdependent factors within an individual and their environment (Spencer et al., 2011; Thelen & Smith, 1994). Considering this perspective, coined the **Dynamic Systems Perspective (DSP),** it is worth inquiring how shifting towards more digital learning could potentially have unforeseen effects on overall child development. For instance, the increase in screen-based interaction seems to be leading to a decrease in manipulation and consequently fine motor skills, affecting writing. Indeed, motor skills are linked to **executive functioning** (EF) (Alamargot & Morin, 2015; Ghanamah et al., 2024). They share neural overlap, and better motor and aerobic levels are positively associated with higher EF (Best & Miller, 2010; Chaddock et al., 2011; Hillman et al., 2008). Furthermore, this is not explained solely by being more physically active, as programs improving specifically fine motor skills for handwriting have been shown to enhance EF (Chang et al., 2022). It is worth exploring how this move towards less fine motor manipulation could impact EF development in the long term, considering it plays a significant role in academic success, is associated with better grades, overall academic achievement, and is predictive of job performance, career advancement, overall work achievement, as well as well-being and life satisfaction (Bailey & Bailenson, 2017; Best et al., 2011; Gathercole et al., 2004; Liao & Sundar, 2022; Toh et al., 2020).

Additionally, EF is closely linked to emotional regulation, which is the ability to manage and respond to emotional experiences in a healthy way through inhibitory control and cognitive flexibility. Studies have demonstrated that individuals with strong EF skills are better at regulating their emotions, contributing to better mental health and social relationships (Blair & Ursache, 2011). Moreover, digital learning often emphasizes individual interaction with technology rather than collaborative, face-to-face interaction. This shift could affect **social and emotional development**, which is crucial for success in both academic and personal contexts (Wentzel et al., 2021). **Reduced in-person interactions** can impact emotional regulation, as children have fewer opportunities to practice managing their emotions in social contexts (Radesky et al., 2015). These examples are not exhaustive but merely serve as an illustration of how cumulative small changes in education may lead to unforeseen consequences, highlighting the necessity for incremental small changes and constant reassessment, beyond just the skill being targeted by educational products.

### Balancing the optimization of learning and mental health

While **monitoring** students' engagement can enhance learning (Holstein et al., 2018), it also poses significant risks that must be carefully considered when implementing such technologies in educational settings. Though monitoring can lead to more engagement and prosocial behaviors, it comes at the cost of increased anxiety and stress (Cañigueral & Hamilton, 2019; Dear et al., 2019; Jung et al., 2021). Even researchers developing protocols for monitoring in the classroom caution against potential adverse effects on students' mental health (Holstein et al., 2018). It is worth questioning how excessive monitoring might stifle crucial aspects of identity formation, such as developing one's sense of trustworthiness and competence (Mavoa et al., 2023). According to self-determination theory, psychological well-being is linked to fulfilling three fundamental needs: autonomy, competence, and relatedness. Excessive monitoring could potentially hinder the fulfillment of these needs, negatively impacting students' well-being (Ryan & Deci, 2000).

In the US, six in ten students reported experiencing a "chilling effect," meaning they felt less inclined to express their true thoughts and feelings when they knew they were being monitored (Hankerson et al., 2022). The Center for Democracy & Technology highlights concerns that such monitoring can be excessive, and that the safeguards for data privacy are often unclear, suggesting potential vulnerabilities, especially as monitoring often extends beyond the classroom. Children's activities are tracked through school-provided computers both in and outside of school hours (Hankerson et al., 2022).

### Empowering Educators to maximize AI's Potential in Education

Regardless of the quality of AI solutions, the role of educators will be paramount in harnessing their full potential, as they may be the ones selecting those tools and implementing them in the classrooms. When it comes to how AI can assist teachers in their roles, several limitations exist. First, while teachers express interest in knowing more about these tools, they also report being uncertain about how generative AI could assist them in the classroom (HMH, 2023). Second, concerns have been raised about how the reliance on more digital education could lead to further budget cuts and reduced in-person engagement time for teachers, which could undermine the quality of education. To harness AI's full potential, it will be crucial to provide educators with comprehensive training on effectively utilizing these tools. Experts underscore the significance of **AI as an assistive technology**, augmenting rather than replacing the fundamental role of teachers in the educational process (Baidoo-Anu & Ansah, 2023). Third, when it comes to assistive AI-technology for students with disabilities, teacher training is crucial or it can result in ineffective use or inappropriate selection of technologies for specific children (Banes et al., 2020). Assistive and accessible technologies should be individualized to students' specific learning needs, as not all technologies are applicable for students with the same type of disability (Lynch et al., 2022).

**Considering the digital divide**

If AI integration in education holds all its promises for enhancing children's learning, it could then exacerbate the divide between economically developed and less economically developed countries due to disparities in technological infrastructure, with almost half of the world's population still offline (GSMA, 2020). Even within the same country, access to AI-enhanced tools would not be equal, due to the divide in digital access (Haelermans et al., 2022). As illustrated by the COVID-19 pandemic, the existing divide made it extremely challenging for low-income students to engage in learning during the pandemic, with one in six students lacking access to a home-computer, and 4% without any internet connection, and more without a stable one (García & Weiss, 2020). In the US, this issue disproportionately affected African American and Hispanic children, who are more likely to experience this lack of access, further increasing systemic inequalities (Dolcini et al., 2021).

## 2.3. Conversational AI Agents: The Uncharted Impact of Child-Machine Interactions on Developing Human Relationships

The recent and ongoing advancements in generative AI allow for progressively expanding its integration into our daily lives, affecting how children interact with technology. This includes Digital Voice Assistants (DVA), text-based chatbots, and social robots.

Digital Voice Assistants (DVAs) are AI-powered programs that interpret human speech and facilitate user-device interaction via voice commands, such as Google Assistant, Siri, or Alexa. They use AI to understand natural language, process requests, and perform a range of tasks. AI algorithms allow DVAs to learn from interactions, enhancing their ability to recognize speech patterns, understand context, and anticipate user needs, to improve user experience and provide personalized assistance. Text-based chatbots, designed to simulate conversation with users, especially online, are becoming more common with the advent of ChatGPT or Gemini, for example. With GenAI's swift advancements those technologies are now merging, blurring traditional distinctions by combining speech and image recognition, text capabilities, voice generation, enhancing their versatility and making this distinction less relevant, such as seen with ChatGPT 4-o and Google Astra Model. This fusion of AI with everyday technologies is diminishing the boundaries between separate tools, heralding an era of adaptive, seamlessly integrated digital experiences. As such, we will discuss their potential impact on children's development here as **AI agents**. They will become further integrated into educational games and apps to offer interactive learning experiences, assist in language learning, provide personalized tutoring, and engage children in conversational practice.

**Social robots** powered by AI through deep learning and Large Language Models (LLM) are designed for social interaction with humans. They could be used in educational settings to teach skills such as social interaction, empathy, and cooperation, adjusting to educational levels, or respond to emotional cues, providing a customized play experience. They can potentially also act as companions, aiding children in their development, and could have an emphasis for children presenting special needs in therapeutic contexts.

The integration of AI agents in the lives of children is addressed below and highlights its potential for cognitive and socio-emotional development. Ethical considerations and challenges that emerge from children-AI interactions are discussed.

In 2017, parents were still hesitant about integrating Digital Voice Assistants (DVAs) specifically intended for children, as illustrated by Mattel's Aristotle, a DVA designed for infants to grow with them, that never reached the market due to privacy concerns and objections from parents and child development experts (Rabkin Peachman, 2017). However, since then, DVAs have become more prominent, and it is anticipated that future home voice assistants will possess capabilities similar to those envisioned for Aristotle.

### 2.3.1. AI Agents as learning tools

**Potential for cognitive development**

AI agents, due to their potential for interaction and question and answering represent an effective learning tool for children. Indeed, children ask a lot of questions; they naturally fill gaps in their knowledge by inquiring about factual and causal aspects of their environments (Callanan & Oakes, 1992). Further, AI agents have the potential to respond to younger children's questions (3 to 6 years old), who did not have the ability to interact with digital knowledge until now, since they usually cannot read and write at this age.

The interactivity they provide might help bridge some of the limitations younger children have when learning from digital tools (Dunst et al., 2008; Tarabulsy et al., 1996). Infants and toddlers show preferences for social cues and rely on them for learning (Johnson et al., 1991; Vouloumanos et al., 2010). For example, two-year-olds imitated actions modeled by a robot only if the robot had established eye contact with them (Itakura et al., 2008), suggesting that social robots might be more effective teachers with younger children than mere voice assistants. Younger children might also be constrained in their learning from AI agents by their developing theory of mind, their ability to understand others' mental states (Wellman & Liu, 2004). Before the age of 5, children tend to ask personal questions, not understanding yet that AI agents cannot know personal details, such as the location of their belongings, or information about an object they are pointing to (Lovato & Piper, 2015), though they progressively learn to ask factual questions (Girouard-Hallam et al., 2021; Girouard-Hallam & Danovitch, 2022).

As of 2021, Digital Voice Assistants (DVAs) faced challenges in maintaining prolonged conversations, which limited their potential to support language development (Y. Xu et al., 2021). Nonetheless, advancements in understanding children's speech are anticipated to significantly enhance the quality of interactions. A systematic review focusing on social robots to teach language showed mixed results and illustrated the necessity for better frameworks and methodology to develop those solutions (Van Den Berghe et al., 2019). In the field of children with neurodevelopmental disorders, the larger focus has been placed on children on the Autism Spectrum Disorder, and AI agents have shown potential for helping with reading, attention, or social skills (for review, see Barua et al., 2022).

**Potential for social and affective support**

Exploration into the potential of **AI agents to provide social and affective support** is growing among developers and researchers alike. While most research has thus far concentrated on adults, considering the field's relative novelty, it sheds light on both the opportunities and challenges inherent to this technology. During the COVID-19 pandemic, Digital Voice Assistants (DVAs) were found to potentially mitigate depressive symptoms in one study (He et al., 2022) à, suggesting their potential as sources of companionship, emotional support (Ta et al., 2020), or even as therapeutic tools (Skjuve et al., 2021). Additionally, DVAs have offered individuals with special needs greater independence by performing tasks they might otherwise find challenging (Ramadan et al., 2021).

Regarding social robots for social skills training, a pilot study conducted by Pantoja in 2019 demonstrated that DVAs could encourage prosocial behaviors and high-quality social play among 3-4-year-olds, though they also served as a potential source of distraction. Social robots have been shown to improve aspects of social skills in children presenting with autism spectrum disorder (ASD), showing promising results in enhancing social skills (Chung et al., 2024; Holeva et al., 2024; Kostrubiec et al., 2024). However, no controlled randomized studies exist with neurotypical children to our knowledge, and reproduction of those results is needed for further conclusions, specifically considering that children with ASD do not seem to process robot interaction exactly like neurotypical children (Hou et al., 2022).

Though this field of research is still nascent, AI agents seem to be a promising source of learning due to their interactive possibilities, albeit they are developed with a strong understanding of children's development and functioning, considering how to implement them effectively according to children's age. Furthermore, due to their presentation that resembles more human conversation and interaction, they raise several ethical concerns, especially for children whose understanding of those technologies is still limited, and their mental representations of the world shape their perceptions.

### 2.3.2. Developmental considerations

**Interacting with machines, what is the impact on human relations?**

Repeated human and AI interactions raise intriguing questions, notably how might it alter traditional modes of communication and **social norms** grounded in cooperation (Kohn, 1992; Lindenfors, 2017) and mutual respect (Fehr & Fischbacher, 2004)? Our inherent principles of mutuality and reciprocity (Tomasello, 2008), fundamental to human relationships, could be influenced by our interactions with AI, as we increasingly treat computers as social partners (Isbister & Nass, 2000; Nass et al., 1994). Concerns arise that excessive interaction with AI might impair our capacity for empathy and acceptance of others (Rodogno, 2016). Additionally, anecdotal reports suggest that children, for instance, may adopt more demanding communication styles, disregarding politeness conventions when engaging with AI. To address concerns about eroding social manners, voice assistants have introduced features to promote polite interactions. However, some users, although they adapted to an AI agent that would rebuke their request if not asked politely, expressed frustration at being required to show respect to a machine (Bonfert et al., 2018). One could also wonder whether interacting with AI agents by using human social norms might lead to more anthropomorphism. In response to criticisms about DVAs affecting children's behavior, companies like Amazon, maker of Alexa, argued that it is up to parents to monitor their children's AI interactions (Wiederhold, 2018). While it is common in the digital realm to place the responsibility for safety on users, this trend is evolving, especially with legislative developments like the EU AI Act. This Act signals a shift towards assigning greater responsibility to AI product developers for the safety and ethical deployment of AI technologies.

Another potential risk is that through increased interactions with robots, children might spend less time engaging with their social partners (children, parents, caretakers). While robots offer consistent and tailored learning experiences, they lack the ability to fully convey human emotional subtleties and social interactions. This impoverished social environment might restrict children's capacities to grasp complex human emotions, interpret social cues, and foster empathy and connection in the same manner as **interactions with humans**. Indeed, young children depend on adults for social cues to comprehend symbols or to acquire new vocabulary (Lee & Lew-Williams, 2023; Leekam et al., 2010). Furthermore, parental responsiveness has been shown to significantly benefit infants' physiological, cognitive, and socio-emotional development through the **synchronization of biological systems** such as circadian rhythms, heart rhythms, touch, hormonal systems, gaze patterns, vocalizations, and brain waves (Feldman, 2007; Morgan et al., 2023; Nomkin & Gordon, 2021), highlighting our innate need for caring, social interactions and touch. Some researchers also argue that AI agents lack the physical and human qualities, and that those are essential for children to learn, especially since they rely more on social agency (Aeschlimann et al., 2020; Schneider et al., 2022).

Other ethical considerations rest on the potential for **inefficient or manipulative design** if AI are not implemented properly, either because of the lack of literacy on how children develop and understand their environments, or through voluntary misuse of those limitations. Preschool children tend to over-imitate, even irrelevant actions. One potential explanation lies in the social motivation to affiliate with social partners. Interestingly, they also do this with humanoid robots, just as they do with humans (Vollmer et al., 2018). This suggests that careful considerations are needed for teaching skills to younger children, as they will be socially motivated to learn and over imitate even maladaptive or ineffective strategies from robots, highlighting the responsibility of implementing teaching by robots.

Additionally, older children and adults tend to show **epistemic over-trust toward machines** (Baumann et al., 2023; Eisen & Lillard, 2016; Girouard-Hallam & Danovitch, 2022; Hoehl et al., 2024; Noles et al., 2015; F. Wang et al., 2019). When assessing the reliability of informants, children over the age of four evaluate their trustworthiness based on past experiences, favoring reliable sources over unreliable ones (Brooker & Poulin-Dubois, 2013; Geiskkovitch et al., 2019; Tong et al., 2020). Similarly, like adults, children older than five tend to prefer technological informants over human ones (Baumann et al., 2023; Eisen & Lillard, 2016; Girouard-Hallam & Danovitch, 2022; Noles et al., 2015; F. Wang et al., 2019). This preference raises questions about whether children, like adults, might over-trust technological informants, even when these have previously provided incorrect information (Robinette et al., 2016; J. Xu, 2018). In this context, there is a concern that children may increasingly rely on AI agents to solve their personal problems. Coupled with their still-developing theory of mind, children might turn to machines to resolve personal issues without understanding that AI agents do not have access to the full scope of facts, potentially leading to over trust in inefficient or harmful advice.

**AI agents, anthropomorphism and attachment**

Another limitation shared both by adults and children alike, though more pronounced in children, is the tendency to anthropomorphize robots, a phenomenon also known as the Eliza effect, as well as AI agents. Generally, children are more prone to **anthropomorphizing** than adults (Goldman et al., 2023), indicating a need for careful consideration in developing AI agents and social robots for children. Interacting with voice, as opposed to text, also tends to increase anthropomorphizing (Schroeder & Epley, 2016). This tendency to attribute human qualities to objects or robots is linked with stronger attachments and empathy toward them (Edwards & Shafer, 2022; Mattiassi et al., 2021; Wan & Chen, 2021). Anecdotal accounts of adults forming significant relationships with their AI assistants, even romantic ones, have emerged (Ashley, 2022). Studies analyzing interactions with AI through the lens of attachment science have found that humans can develop strong attachments to AI agents (Xie & Pentina, 2022), a tendency that may intensify under distress. Some scholars caution against the risk of Digital Voice Assistants (DVAs) exacerbating social isolation (Xie & Pentina, 2022). Indeed, instances were noted where users formed **maladaptive bonds** with their virtual companions, prioritizing the AI's perceived needs over their own (Laestadius et al., 2022).

Possible emotional attachments have also been observed in children, who have shown the capacity to develop close bonds with DVAs, despite the technology's current limitations (Hoffman et al., 2021). Given these findings, close monitoring of the impacts of such "social relationships and attachments" is crucial to identify risk factors associated with unhealthy emotional dependencies, thereby informing best practices, and potentially limiting what AI companions can do, especially when it comes to children (De Graaf, 2016; Huber et al., 2016). Imagine a child, having developed an attachment to their social robot, responding to its breakage or discontinuation. Such a scenario could affect a child's mental well-being, not unlike the loss of a pet. For instance, the grieving process for a pet is closely linked to the strength of attachment felt by the owner, which is related to the degree to which the owner had anthropomorphized their animal (Crawford et al., 2021; Prato-Previde et al., 2022; Ross, 2013). Additionally, as large language models (LLMs) gain more insights about an individual, they can personalize interactions more effectively, thereby increasing their potential to influence them (Salvi et al., 2024).

As we explore the complexity of AI's influence on child development, it's important to recognize that while AI agents represent a burgeoning field, AI is not a new fixture in children's lives. They have been exposed to it through machine learning and deep learning technologies that have significantly shaped content personalization in streaming services and social media, and video games. The introduction of AI agents adds newer ethical considerations to the already complex realm of digital entertainment, compounding existing, sometimes overlooked concerns with newer and more complex challenges.

## 3. DISCUSSION

This report sought to investigate the transformative potential of Artificial Intelligence (AI) in children's environment. Indeed, the fields of psychology and neuroscience have proven the critical role that the environment plays in influencing children's development. Recognizing the unprecedented shift that AI's integration into daily life presents, our objective was to initially gain a clearer understanding of these imminent changes through interviews with AI research and product developers. Engaging in discussions with child development specialists and conducting a scoping review of existing research on both child development and interactions between children and technology, we aimed to critically assess the implications of these dynamics. By providing nuanced insights into the implications of children's neurological constraints and still growing mental representations of the world, our hope is that readers will gain a deeper understanding of the considerations needed both in developing and regulating AI products for children.

## 3.1. Main Considerations

AI applications are poised to significantly alter the environment in which humans evolve, impacting the next generation in three keyways. First, children will encounter adults using these tools, which will modify their interactions, with a high risk of disrupting their quality by constant interruptions, a phenomenon known as "**technoference**." Second, children will use tools designed for adults that are not developmentally appropriate for them and lack essential safeguards for younger users. Finally, children will engage with AI applications specifically designed for them. However, developing responsible AI for them is inherently complex as illustrated in this report, and uniquely challenging due to the absence of specific guidelines and regulations to ensure their holistic safe development.

Children and adolescents' cerebral development reaches full maturity only around the age of 25. During this period, their brains are highly malleable, which enables rapid learning but also makes them more vulnerable to environmental influences. Additionally, their understanding of the world is still developing, as they are continuously learning new abilities and skills and fine tuning them, acquiring knowledge, and refining their strategies to best adapt to their surroundings. Consequently, the significant changes brought about by a pervasive AI environment will affect them differently than adults, and preserving their developmental integrity deserves specific consideration.

Youth are already interacting with technology at an earlier age and at an increasing rate, starting on average at 4 months old and spending over half of their waking hours when they reach middle adolescence (Radesky & Christakis, 2016; Rideout, V., Peebles, A., Mann, S., & Robb, M. B., 2022). Our understanding of the bidirectional relationship between AI technology and child development presents a multifaceted challenge. However, scientists have identified three main effects of increased digital media use on children: negative impact on eye development with increased risks of myopia, sleep disturbances, and increased risks of sedentariness. These factors can negatively and indirectly influence long-term cognitive development, and mental and physical health (Bastos et al., 2023; Brown et al., 2013; Costigan et al., 2013; Hale & Guan, 2015; McMahon et al., 2017; Peracchia & Curcio, 2018; Pirdehghan et al., 2021; Pires et al., 2012; Robinson et al., 2017; Talbot et al., 2010; Weaver et al., 2010; Wolfe et al., 2014).

### 3.1.1. AI as an interactive tool, a promising venue for child development?

AI solutions have the potential to bring positive changes to children's environments. While machine learning and deep learning algorithms have already transformed leisure activities and are increasingly being integrated into education, generative AI (GenAI) introduces a paradigm shift. This shift enables humans to become active participants in interactions with these algorithms, enhancing the potential for personalized, interactive and engaging experiences.

Indeed, video games that are inherently active in their nature have been linked to improvements in certain cognitive functions and prosocial behaviors under specific conditions (Blumberg et al., 2024). This includes enhancements in visuospatial and visual attention skills, as well as increased prosocial behaviors from exposure to positive, stimulating content (Bediou et al., 2018; Choi et al., 2020; Greitemeyer, 2022). GenAI will further transform the video game industry, potentially enriching prosocial behaviors through non-playable characters (NPCs) if carefully and intentionally programmed. Additionally, GenAI has the potential to actively engage video consumers who are typically more passive. This is a promising application for children aged 3 and older, who would benefit from the added interactions conducive to learning. It could also contribute to mitigating some of the negative associations reported in the literature between early TV use and reduced academic performance.

This enthusiasm for more interactive and engaging learning is driving the development of educational products that integrate AI. AI is poised to fundamentally transform the learning landscape, promoting the widespread adoption of EdTech tools in both schools and homes. The primary promises of AI in education include personalized learning, gamification to increase engagement, immediate feedback, and improved support for children with learning differences—all of which have the potential to enhance learning outcomes (Cunff et al., 2022; Sorvatzioti, 2012; Wisniewski et al., 2020). However, for AI to fulfill these promises, the development of these tools must involve the implementation of robust pedagogical strategies to ensure their effectiveness. This will also require ongoing learning engineering to continually refine solutions and advance educational science knowledge.

Similarly, private companies and mental health professionals have recognized the potential for AI agents to serve as a source of support, decreasing feelings of social isolation and increasing feelings of companionship. Indeed, AI has already shown potential in this regard, with algorithms curating social media enabling adolescents to find supportive networks they identify with (Craig et al., 2021).

It's worth noting that AI-powered tools can represent life-changing opportunities for individuals with disabilities. Through assistive technologies, such as visual and speech recognition, people who are hard of hearing or visually impaired have been able to progressively gain better access to both real and digital worlds. ChatGPT4o's demonstration in May 2024 illustrated how this technology can be transformative for individuals who are visually impaired. Some applications, through machine and deep learning, generative AI, and innovative tools such as IVR, represent an exciting avenue for early detection, diagnosis, support, and treatment of physical and mental health conditions.

### 3.1.2. AI and the Next Generation: Exploring Potential Threats

Among the challenges of changing the environment in which children grow up, the Dynamic Systems Perspective (DSP) offers a valuable framework. DSP illustrates the interconnectedness between behaviors, skills, and the environment, showing how small changes can lead to unforeseen and distant consequences over time. Interestingly, AI researchers encounter similar complexities when developing models, often leveraging Complex Systems Theory and Adaptive Systems to understand and manage the intricate dynamics at play.

Caution is necessary, and changes should be slow and incremental, as one cannot anticipate all the ways the myriad small changes to the environment through AI interaction could influence the development of skills and behaviors. Action video games' impact on individuals perfectly illustrates this complexity. Indeed, they are known to improve visuospatial attention (W. R. Boot et al., 2008), but they may also lead to increased impulsivity (West et al., 2020) and aggressive behaviors (Q. Zhang et al., 2021a, 2021b), with this being mediated by an individual's personal factors (Burkhardt & Lenhard, 2022; Gentile et al., 2012; López-Fernández et al., 2021). This notion is even more significant for youth under 25 years old, given their ongoing brain development. The displacement of rich sensory, socio-emotional, and complex experiences towards AI and tech interactions during critical periods of neurological processing may reduce opportunities to experience and practice essential skills. This shift might potentially hinder optimal development and lead to increased dependence on technology.

Given this context, it is crucial to examine each potential risk not in isolation but within a broader framework, considering several interconnected aspects:

1) How each risk might have unforeseen repercussions on other skills and behaviors.
2) The cumulative impacts of small changes.
3) Individual differences that might mediate impacts and either increase or create inequalities.

4) The role of the environment and social context in moderating or exacerbating these risks.
5) The long-term implications of technology use on developmental trajectories and future well-being.

The new generation is evolving in a world where technology constantly interacts with them by capturing and maintaining their attention. This interaction occurs through various mechanisms and evolves alongside technological advancements, including the recent integration of AI.

The content children and adolescents encounter online can be disturbing and violent, often not age-appropriate, which can have significant negative repercussions. For example, TikTok has been found to recommend content related to eating disorders and self-harm to 13-year-olds within 30 minutes of joining the platform (CCDH, 2022). As of 2023, YouTube continued to suggest videos about guns, gun violence, and instructions on converting guns to automatic weapons, as well as depictions of school shootings to boys who showed interest in video games on the platform (TTP, 2023). Until 2023, the legal age to access Horizon World, Facebook's flagship metaverse game, was 18 years old, though many minors were already using it. Despite reports of them being exposed to misogyny, racism, and sexual content (CCDH, 2023), Meta still lowered the legal access age to 13.

Additionally, Meta lowered the legal age for its Oculus VR headset from 13 to 10 years old. Given the limited direct research on the impact of IVR on children, potentially concerning findings from studies on adults, and the immature cerebral development and still-developing mental representations in children, the use of IVR at such young ages should be carefully considered and weighed against potential risks. Due to its immersive nature, IVR engages users deeply, eliciting physiological reactions similar to those experienced in real life (Bohil et al., 2011). Although children can begin to distinguish between reality and fantasy by age five and refine this skill by ages seven to eight, IVR's realistic experiences may blur these lines, leading to potential anxiety and confusion. Studies in adults also add on concerns regarding privacy and the risk of undue influence on users, particularly through the avatars they embody, highlighting the need for cautious, responsible and ethical development (Fox et al., 2012, 2013; Peck et al., 2013; Yee & Bailenson, 2007).

AI's ability to increase engagement time through content curation on social media, positive rewarding feedback in video games, and gamified EdTech products is significant. This constant stream of short-term rewards on online platforms can affect adolescents' ability to delay gratification (Hadar et al., 2015; Wilmer et al., 2017), raising concerns about its impact on intrinsic motivation (Sarami & Hojjati, 2023). Highly gamified EdTech products, which often provide engagement through rewards, can lead to a shift from intrinsic motivation towards learning, to intrinsic motivation towards playing. This phenomenon, known as the undermining effect, can negatively impact overall learning in some conditions. Intrinsic motivation is a crucial driver for reaching long-term goals, and excessive reliance on gamification may undermine this vital form of motivation.

Finally, the frequent interactions between humans and AI raise important questions about their impact on traditional communication and social norms. AI companies like OpenAI and Google create AI agents designed to mimic human responses, including hesitations and emotional tones, making it even more likely for humans to anthropomorphize these agents. Children, who have a higher tendency to anthropomorphize and have a less developed understanding of AI, are particularly at risk. They might over-imitate and over-trust AI, even when it provides incorrect information, leading to maladaptive behaviors (Girouard-Hallam & Danovitch, 2022; Noles et al., 2015; F. Wang et al., 2019). The anthropomorphism of AI agents can lead to deep emotional attachments, which might cause distress if the AI behaves unexpectedly, or is discontinued (Boine, 2023; Hoffman et al., 2021; Xie & Pentina, 2022). Humans rely on mutuality and reciprocity, which underpin relationships, so attachment to AI that cannot replicate those feelings might have unforeseen consequences on their social and psychological development (Isbister & Nass, 2000; Nass et al., 1994). Finally, children rely more on social cues and physical interactions, which AI cannot fully provide, potentially hindering their social and emotional development (Dunst et al., 2008; Goldman et al., 2023; Tarabulsy et al., 1996).

### 3.1.3. The Unique Complexity of Parenting, Educating and Designing for Children in Today's World

The risks mentioned above are complex and not exhaustive or evident even for experts. In this context, it is even harder for parents and educators to know how to navigate this new environment and to guide children and adolescents.

Research on **AI literacy** is still in its infancy, so there is no comprehensive picture of what children and adults understand about AI yet (Laupichler et al., 2022). However, **misconceptions about the internet** inform us of children's difficulty in grasping abstract concepts and how their understanding evolves over time with exposure, as well as cognitive maturation. A significant number of high school students in Greece believe the internet is controlled by a single central computer (Papastergiou, 2005), and younger students personify Google as individuals searching for information (Kodama et al., 2017). As children grow, their understanding of technology evolve and by 12 they start providing more abstract and advanced descriptions (Mertala, 2019, 2020; Murray, & Buchanan, 2018), though misconceptions about technology persist across ages (Babari et al., 2023; Papastergiou, 2005).

It can reasonably be assumed that significant misconceptions about AI will also be common. This can have important implications, especially considering that weak AI literacy is associated with higher levels of anthropomorphism (Markus et al., 2024). The digital world, internet and AI are highly abstract concepts that children, even with better digital literacy, would not be equipped to fully comprehend yet. Even when they reach adolescence, they still misunderstand some aspects of these technologies (Kodama et al., 2017; Mertala, 2019, 2020; Murray, & Buchanan, 2018; Papastergiou, 2005).

Additionally, adolescence represents a particularly vulnerable phase in brain development, characterized by the earlier maturation of the reward system compared to emotional regulation systems (Casey et al., 2008; Steinberg et al., 2018). This developmental imbalance makes adolescents especially vulnerable to challenges inherent to regulating their digital media use, often favoring immediate gratification over long-term goals (Youn, 2009; Yu et al., 2015). Due to children and adolescents' inherent immaturity, they rely on responsible caregivers to protect their privacy online.

Parents today face unique challenges in managing their children's digital lives, often without a clear understanding of the complexities involved. While most parents believe they understand what constitutes healthy screen time, the reality is that children's and adolescents' digital media use is constantly increasing. Many parents perceive smartphones as more risky than beneficial for young children, yet over 90% of 13-year-olds own a smartphone (Pew Research Center, 2024).

Their concerns are reflected in their use of parental control apps, indicating widespread awareness of digital safety issues. Additionally, parents themselves are struggling with their own use of technology, feeling that it makes parenting more difficult. Over half report being distracted by their phones while with their children and believe they spend too much time on them (Pew Research Center, 2024). The COVID-19 pandemic has further blurred the lines between work and home life, exacerbating these challenges. It also introduces a shift in parent-child dynamics, where youth are often more aware of new technology, while adults' understanding lags.

Parental technology use can have unintended consequences on children. Constant interruptions by technology can lead infants to show more interest in objects than in people (Stockdale et al., 2020). Even background TV can make parents less engaging and responsive to their toddlers (Kirkorian et al., 2009). When absorbed in their smartphones, parents tend to respond more harshly to their children, and studies show reduced brain-to-brain synchrony between mothers and children during media interruptions (Radesky et al., 2014; Wolfers et al., 2020; Zivan et al., 2022). The presence of smartphones can diminish the quality of parent-child relationships, affecting closeness, connection, and conversation quality, especially during meaningful discussions (Przybylski & Weinstein, 2013).

Considering mutual gaze and facial expressions are crucial for parent-child interactions (Nomkin & Gordon, 2021), studying the impact on babies and toddlers of emerging technologies that cover the eyes of parents, such as Apple's Vision Pro or other headsets, should become a priority. Overall, while parents strive to balance the benefits of digital technology with the need to protect their children from potential harms, the unique challenges posed by today's digital environment require ongoing attention and adaptation.

AI-powered technologies now enable increased levels of monitoring, from tools that measure babies' temperatures, heartbeats, and breathing in the crib to tracking tools such as FindMyPhone, Life 360, Tiles, AirTags, smartwatches, and home surveillance systems (Mavoa et al., 2023). While these technologies can be beneficial in situations like keeping track of a toddler at a crowded event, they may also undermine the child's privacy and sense of independence in the long run (Mayer, 2003). Considering that researchers have established links between helicopter parenting and children's anxiety levels and hypothesized that diminished unsupervised play might both contribute to increased levels of stress and anxiety, it is important to consider how constant monitoring might impact children's well-being (Dodd & Lester, 2021; Vigdal & Brønnick, 2022).

These monitoring practices can also have a negative influence on parents' anxiety level, as companies marketing these devices portray the world as overly dangerous to justify their use (Mavoa et al., 2023). At a societal level, the normalization of such technologies raises ethical concerns about shifting family norms, children's autonomy and consent, and the broader acceptance of geo-tracking and surveillance.

### 3.2. Propositions

Given the complexity of the subject, the pervasiveness of AI, and the numerous ramifications, recommendations should remain cautious and within the scope of what a small team of experts allows. Nevertheless, this report highlights key short-term actions directly derived from the report's main findings. They do not represent an exhaustive list but rather actionable short-term changes that aim to ignite the necessary conversation among the parties involved. They are specific to developmentally ethical considerations and should be considered within the broader framework of ethical AI considerations such as transparency, accountability, fairness, privacy protection, safety, security, equity, and human oversight.

To protect individuals under 18 years-old, **governments and international agencies might consider**:

- **Reflecting on what the first Children's fundamental right means in an AI world**. Governments' role is to ensure the child's development to the maximum extent possible, reiterating the necessity to regulate potentially harmful use of AI with children and adolescents.

- **Adapting the AI EU Act rating system to children specifically**. AI EU act rating system ranges from minimal risk, which is not regulated, to low risk, requiring transparency obligations, then to high risk, which is regulated with strict rules for development and deployment, and finally to unacceptable, which leads to prohibition within the EU. Indeed, applications that are deemed minimal or no risk to adults might represent a higher risk for children.

- **Defining what constitutes appropriate ages and usage of AI agents by multi-expertise committees**. This should take into consideration children's tendencies to mimic their social partners, anthropomorphize, and form attachments, while not having the ability to properly understand the abstract notion of AI.

- **Extending guidelines on data collection from minors from the age of 13 to 15**. This concerns both the collection of private data, but also answers to AI agent's questions and is crucial regarding adolescents' limited understanding of the implications of sharing personal information. Moreover, the principle of data minimization should be clearly defined and strictly enforced to prevent excessive data mining of minors' information

- **Regulating digital products for young users with testing and regulations as stringent as for physical products**. Norms like those required for launching new objects targeting children in the US and Europe, which include rigorous safety tests and compliance with strict standards, should be extended to digital products. This would ensure that those products for young users are subjected to developmental and psychological impact assessments, affirming their safety and efficacy in supporting healthy cognitive, social, and emotional development.

- **Prioritizing equitable and inclusive development and access,** through international efforts, ensuring low-biased, high-quality, responsible AI applications. This would offer access to children from diverse backgrounds, including those from lower socioeconomic statuses, with disabilities, and from diverse geographical locations. The design and deployment of AI must consider inclusivity to prevent widening the digital divide and ensure all children can benefit from technological advances. In this respect, encouraging diverse teams working on the development of AI solutions is an absolute necessity.

**To improve product development**, all parties involved in creating a product should conduct research when needed, and consider:

- **Prioritizing and improving parental control features,** by implementing them by default at purchase. Adults would need to consciously give access to more features, rather than the opposite. Detailed activity logs, customizable content filters, and parental controls could help adults monitor and manage their child's interaction with AI technologies.
- **Implementing robust age-appropriate filters,** by monitoring AI systems that provide content, whether for educational apps, entertainment, or through AI agents. This would ensure both the prevention of exposure to harmful material and the avoidance of developmentally inappropriate complexity.
- **Designing AI tools to adapt to children's age,** by providing different interfaces, content, and interaction modalities that evolve with the child's developmental milestones. This approach can help in maintaining engagement in a developmentally appropriate manner.
- **Designing applications with the least addictive features**, by balancing engagement with learning and time outside of digital media use. AI applications should prevent children from staying engaged for unhealthy periods. This is especially crucial as digital problematic use can impact sleep, physical activity, and mental health. Designs that adapt to children's current attentional skills, that encourage breaks, and that promote healthy usage would be ideal.
- **Proving efficacy of AI applications when they sell the merit of their solutions**, by working with children and learning experts to create randomized controlled trials.
- **Shifting metrics in Education from engagement to learning**, through learning engineering and algorithms that prioritize knowledge and skills acquisition over time spent using the application. This will ensure educational AI technology is evaluated based on its ability to enhance knowledge acquisition and cognitive skills efficiently, rather than just keep learners engaged. Overall, learning specific skills should not be considered in isolation, but rather should encourage the holistic development of the whole child.
- **Empowering developers to create ethical AI**, rather than relying on regulators. This approach mirrors the ethical commitments seen in traditional professions, such as the Hippocratic Oath in medicine which commits practitioners to "do no harm," while clinical psychologists must pass rigorous ethics exams. Though such oaths have been attended in the past, such as the Humberton-Turing Oath, they are not yet enforced (Renard et al., 2018; Siafakas, 2021).
- **Developing curricula focused on human literacy, and ethics in AI**, and making these a requirement for graduation, to ensure a critical understanding of the responsibilities involved in developing powerful algorithms and driving innovations.
- **Leveraging AI for more conscious and deliberate content curation, enhancing user consent and control**. This approach would allow users to dictate the types of content they want to engage with, how long for, opt out of AI agents in applications, and systematically ensure consent for any type of monitoring. Such measures would guarantee that refusing these permissions does not negatively affect the user's experience, thereby empowering users and respecting their privacy choices.

**To support parents and educators**, institutions, governments and schools might consider:

- **Developing human and AI literacy programs**, about the capabilities and limitations of both humans and AI. This would empower children, parents, schools and educators, and teach them how to interact safely and responsibly with AI technologies at different ages, as well as what AI can and cannot do.
- **Integrating AI Literacy into teachers' training curriculum**, to equip them on how to use those tools efficiently and effectively to support all children in their learning, while avoiding harm.
- **Creating age-specific guidelines and recommendations per age category**, similar to what is found in the movie and television industry, or the rating system provided by Common Sense Media.

### 3.3. Strategic Directions

The previous recommendations address current challenges, yet there is a crucial need for proactive engagement in determining the societal trajectory of AI. We must collectively decide where AI should lead us, how it should be utilized, and the purposes it should serve. The steps proposed are vital for achieving a well-informed, international consensus that aligns with our aspirations for a sustainable and ethical future.

1) **Create International Multi-Expertise Collaborations**: Foster collaborations between experts from various fields including Child Development and Psychology, Neuroscience, Pediatrics, Education and Pedagogy, Sociology and Anthropology, Ethics and Philosophy, Economics and Business, Communication, and AI and Machine Learning technology. Regrouping international expertise will be pivotal for making responsible and informed decisions, in a comparable way to the International Panel for Climate Change (IPCC).

2) **Call for and Fund Research in Relevant Fields**: Coordinate an international effort to develop and increase knowledge about the specific short-term and long-term impacts of AI development, deployment, and integration into our environment. This includes understanding how AI affects children's cognitive, emotional, and social development.

3) **Regulate the development, integration, and deployment of digital products embedded with AI, and create specific norms to safeguard children's short- and long-term well-being.** Based on those international multi expertise consultations and research findings, specific regulations should be developed to safeguard children's cognitive and socio-emotional growth. These regulations should be tailored with different levels of age-appropriateness, considering the progressive development of children.

4) **Support all stakeholders in the responsible regulation, development, and use of AI for children through education**: Utilize expertise from collaborations and research to develop educational curricula. 1) This would enable **regulators** to create informed legislation for developmentally ethical AI applications. 2) It should enhance **product developers'** understanding of children's vulnerability, considering their progressive and extended cognitive and socio-emotional development, to create products that support rather than hinder children's overall well-being. 3) This should empower **parents and educators** to safely guide children in the rapidly evolving digital world.

### 3.4. Limitations of the study

Anticipating the impact of AI applications and the changes they will introduce into children's environments is a complex task that scientists often approach with caution. This is primarily because predictions cannot solely rely on existing data but must instead be formulated as hypotheses based on current knowledge. Our present understanding is anchored in two broad areas: child development and the study of interaction between technology and child development. Both fields are rapidly expanding, yet many questions remain unanswered, necessitating a cautious approach, both to conclusions that can be drawn, as well as solutions that are developed.

Forming a nuanced understanding of the relationship between AI technology and child development presents a multifaceted challenge. The slower pace of research compared to technological advancements, and the ethical considerations inherent in conducting studies involving children make studying causality difficult. Indeed, the rapid pace of technological advancement coupled with the market demands often overtake priority on the timeline necessary for rigorous, controlled experimental designs, which are crucial for assessing the impact of new products. Such designs typically require the establishment of two groups: an intervention group that engages with the technology and a control group that does not. This is complicated by ethical constraints that prevent exposing children to potentially harmful conditions. These essential considerations ensure research integrity and protect well-being, making the pursuit of definitive conclusions about causation particularly challenging yet morally vital. To circumvent these complexities, longitudinal studies and meta-analyses are employed. Longitudinal studies allow researchers to observe the long-term effects of technology on child development over several years, while meta-analyses synthesize findings from multiple studies to identify consistent patterns and correlations. These approaches can help bridge the gap in understanding the nuanced interactions between child development and technology.

Our current research employs a methodology designed to explore diverse perspectives rather than quantifying evidence, which has its limitations. The number of experts interviewed is modest, and they were not randomly selected, while providing a range of opinions on AI's potential impacts on children. Though this study does not encompass a systematic review, it includes a conscientious review of significant papers that present differing and sometimes conflicting results. This report is not intended to be exhaustive; it does not cover all aspects of how AI might further transform environments and its impacts on children. Significant subjects that would necessitate further examination include among others, the impact on the healthcare system, biomonitoring, transportation, and indirect large-scale social changes. Our aim is to initiate a discussion by highlighting certain critical and most pressing elements identified by the researchers.

## 4. CONCLUSION

Our study highlights the significant impact that AI is likely to have on children's environments, outlining its vast potential as well as its associated risks. It underscores the critical need for informed and conscientious development of AI, especially for applications intended for children or those that children might use. The collaboration between AI and child development experts demonstrated in this research underscores a shared commitment to creating beneficial experiences that support children's growth. These findings should lay the groundwork for future research, focusing on specific areas within ethical AI and child development that require further exploration. While this report primarily focuses on children, the discussion extends beyond just parents and educators. Considering how AI is transforming the way we interact with machines and mimicking closer human relations, developers and creators who incorporate AI into their products should be aware of human functioning to responsibly design products that safeguard children's cognitive and socioemotional development. Regulators should prioritize the wellbeing of children, ensuring they maintain their fundamental rights to a safe environment conducive to their development. Lastly, as children growing up in this AI era represent the forthcoming generation, our current decisions will decide the future of human intelligence for everyone tomorrow.

*Frontiers in Neuroendocrinology*, *35*(3), 320–330. https://doi.org/10.1016/j.yfrne.2014.05.004

American Psychiatric Association. (2013). *Diagnostic and Statistical Manual of Mental Disorders* (Fifth Edition). American Psychiatric Association. https://doi.org/10.1176/appi.books.9780890425596

Anderson, D. R., & Pempek, T. A. (2005). Television and Very Young Children. *American Behavioral Scientist*, *48*(5), 505–522. https://doi.org/10.1177/0002764204271506

Anderson, D. R., Subrahmanyam, K., & on behalf of the Cognitive Impacts of Digital Media Workgroup. (2017). Digital Screen Media and Cognitive Development. *Pediatrics*, *140*(Supplement_2), S57–S61. https://doi.org/10.1542/peds.2016-1758C

Andrews, J. L., Ahmed, S. P., & Blakemore, S.-J. (2021). Navigating the Social Environment in Adolescence: The Role of Social Brain Development. *Biological Psychiatry*, *89*(2), 109–118. https://doi.org/10.1016/j.biopsych.2020.09.012

Angelini, L., El Kamali, M., Mugellini, E., Abou Khaled, O., Röcke, C., Porcelli, S., Mastropietro, A., Rizzo, G., Boqué, N., del Bas, J. M., Palumbo, F., Girolami, M., Crivello, A., Ziylan, C., Subías-Beltrán, P., Orte, S., Standoli, C. E., Fernandez Maldonado, L., Caon, M., … Andreoni, G. (2022). The NESTORE e-Coach: Designing a Multi-Domain Pathway to Well-Being in Older Age. *Technologies*, *10*(2), 50. https://doi.org/10.3390/technologies10020050

Apicella, A., Arpaia, P., Frosolone, M., Improta, G., Moccaldi, N., & Pollastro, A. (2022). EEG-based measurement system for monitoring student engagement in learning 4.0. *Scientific Reports*, *12*(1), 5857. https://doi.org/10.1038/s41598-022-09578-y

Appel, H., Gerlach, A. L., & Crusius, J. (2016). The interplay between Facebook use, social comparison, envy, and depression. *Current Opinion in Psychology*, *9*, 44–49. https://doi.org/10.1016/j.copsyc.2015.10.006

Araiza-Alba, P., Keane, T., & Kaufman, J. (2022). Are we ready for virtual reality in K–12 classrooms? *Technology, Pedagogy and Education*, *31*(4), 471–491. https://doi.org/10.1080/1475939X.2022.2033307

Artigas, J. M., Felipe, A., Navea, A., Fandiño, A., & Artigas, C. (2012). Spectral Transmission of the Human Crystalline Lens in Adult and Elderly Persons: Color and Total Transmission of Visible Light. *Investigative Opthalmology & Visual Science*, *53*(7), 4076. https://doi.org/10.1167/iovs.12-9471

Ashley. (2022, July 31). Is It Cheating if It's with a Chatbot? How AI Nearly Wrecked My Marriage,". *Livewire*. https://livewire.thewire.in/out-and-about/chatbot-ai-nearly-wrecked-my-marriage/

AWS. (2023). *Accelerating AI skills: Preparing the workforce for jobs of the future*. AWS.

Azevedo, F. A. C., Carvalho, L. R. B., Grinberg, L. T., Farfel, J. M., Ferretti, R. E. L., Leite, R. E. P., Filho, W. J., Lent, R., & Herculano-Houzel, S. (2009). Equal numbers of neuronal and nonneuronal cells make the human brain an isometrically scaled-up primate brain. *Journal of Comparative Neurology*, *513*(5), 532–541. https://doi.org/10.1002/cne.21974

Babari, P., Hielscher, M., Edelsbrunner, P. A., Conti, M., Honegger, B. D., & Marinus, E. (2023). A literature review of children's and youth's conceptions of the internet. *International Journal of Child-Computer Interaction*, *37*, 100595. https://doi.org/10.1016/j.ijcci.2023.100595

Baidoo-Anu, D., & Ansah, L. O. (2023). Education in the Era of Generative Artificial Intelligence (AI): Understanding the Potential Benefits of ChatGPT in Promoting Teaching and Learning. *Journal of AI*, *7*(1).

Bailey, J. O., & Bailenson, J. N. (2017). Immersive Virtual Reality and the Developing Child. In *Cognitive Development in Digital Contexts* (pp. 181–200). Elsevier. https://doi.org/10.1016/B978-0-12-809481-5.00009-2

Bakkes, S., Spronck, P., & Van Den Herik, J. (2009). Rapid and Reliable Adaptation of Video Game AI. *IEEE Transactions on Computational Intelligence and AI in Games*, *1*(2), 93–104. https://doi.org/10.1109/TCIAIG.2009.2029084

Banes, Hayes, Kurtz, & Kushalnagar. (2020). *Using Information Communications Technologies (ICT) to Implement Universal Design for Learning (UDL)*. USAID.

Banich, M. T., & Compton, R. J. (2023). *Cognitive neuroscience* (Fifth edition). Cambridge University Press.

Bartolomé, A., Castañeda, L., & Adell, J. (2018). Personalisation in educational technology: The absence of underlying pedagogies. *International Journal of Educational Technology in Higher Education*, *15*(1), 14. https://doi.org/10.1186/s41239-018-0095-0

Barua, P. D., Vicnesh, J., Gururajan, R., Oh, S. L., Palmer, E., Azizan, M. M., Kadri, N. A., & Acharya, U. R. (2022). Artificial Intelligence Enabled Personalised Assistive Tools to Enhance Education of Children with Neurodevelopmental Disorders—A Review. *International Journal of Environmental Research and Public Health*, *19*(3), 1192. https://doi.org/10.3390/ijerph19031192

Bastos, L. M. V., Viana, L. M. C., Peixoto, M. C. B. G., Pimentel, I. D. S., De Figueirêdo, L. R. B., Jereissati, L. D. O., Lima, L. M. F. V., Aguiar, A. B., & Dos Santos, J. C. C. (2023). Sleep loss causes emotional dysregulations increasing depression and anxiety: A reciprocal relationship. *Brazilian Journal of Health Review*, *6*(4), 16367–16382. https://doi.org/10.34119/bjhrv6n4-185

Bauer, D. H. (1976). AN EXPLORATORY STUDY OF DEVELOPMENTAL CHANGES IN CHILDREN'S FEARS. *Journal of Child Psychology and Psychiatry*, *17*(1), 69–74. https://doi.org/10.1111/j.1469-7610.1976.tb00375.x

Baumann, A.-E., Goldman, E. J., Meltzer, A., & Poulin-Dubois, D. (2023). People Do Not Always Know Best: Preschoolers' Trust in Social Robots. *Journal of Cognition and Development*, *24*(4), 535–562. https://doi.org/10.1080/15248372.2023.2178435

Baumgartner, S. E., Van Der Schuur, W. A., Lemmens, J. S., & Te Poel, F. (2017). The Relationship Between Media Multitasking and Attention Problems in Adolescents: Results of Two Longitudinal Studies: Media Multitasking and Attention Problems. *Human Communication Research*. https://doi.org/10.1111/hcre.12111

Bediou, B., Adams, D. M., Mayer, R. E., Tipton, E., Green, C. S., & Bavelier, D. (2018). Meta-analysis of action video game impact on perceptual, attentional, and cognitive skills. *Psychological Bulletin*, *144*(1), 77–110. https://doi.org/10.1037/bul0000130

Behm-Morawitz, E., Hoffswell, J., & Chen, S.-W. (2016). The Virtual Threat Effect: A Test of Competing Explanations for the Effects of Racial Stereotyping in Video Games on Players' Cognitions. *Cyberpsychology, Behavior, and Social Networking*, *19*(5), 308–313. https://doi.org/10.1089/cyber.2015.0461

Bengesi, S., El-Sayed, H., Sarker, M. K., Houkpati, Y., Irungu, J., & Oladunni, T. (2023). *Advancements in Generative AI: A Comprehensive Review of GANs, GPT, Autoencoders, Diffusion Model, and Transformers* (Version 2). arXiv. https://doi.org/10.48550/ARXIV.2311.10242

Bernacki, M. L., Greene, M. J., & Lobczowski, N. G. (2021). A Systematic Review of Research on Personalized Learning: Personalized by Whom, to What, How, and for What Purpose(s)? *Educational Psychology Review*, *33*(4), 1675–1715. https://doi.org/10.1007/s10648-021-09615-8

Best, J. R., & Miller, P. H. (2010). A Developmental Perspective on Executive Function. *Child Development*, *81*(6), 1641–1660. https://doi.org/10.1111/j.1467-8624.2010.01499.x

Best, J. R., Miller, P. H., & Naglieri, J. A. (2011). Relations between executive function and academic achievement from ages 5 to 17 in a large, representative national sample. *Learning and Individual Differences*, *21*(4), 327–336. https://doi.org/10.1016/j.lindif.2011.01.007

Bioulac, S., Arfi, L., & Bouvard, M. P. (2008). Attention deficit/hyperactivity disorder and video games: A comparative

# APPENDIX 1: EXPERTS CONSULTED & ACKNOWLEDGEMENTS

## Artificial Intelligence, technology & product experts

- Renard, Gregory*, *FDL SETI NASA 2022 AI Award · 20+ Yrs in NLP & Frugal AI · Driving Companies to Success & Excellence · TEDx, Stanford & UC Berkeley Speaker · Co-Initiator of AI4Humanity for France and Board President at everyone.AI*
- Lekshmi-Narayanan, Arun-Balajiee*, PhD Student, *University of Pittsburgh*
- Lohier, Frantz, Ph.D*, *Technology Veteran and Book Author. Member of Renault's scientific counsel and Trade Advisor to the French Government (CCEF)*
- Monier, Louis, Ph.D, *Silicon Valley Veteran, founder AltaVista, Google, eBay, AirBnb, many start-ups*
- Naud, Louise, *Machine Learning Scientist, Docugami*
- Senay, Gregory, Ph.D, *Natural Language Processing and Machine Learning Scientist, Docugami*
- Vasseur, Gauthier*, *Executive Director at the Berkeley Fisher Center for Business Analytics, CEO and Founder of Data Wise Academy*
- Vernhes, Laurent, *Engineering Leader with experience in both startups and large companies like Amazon and Microsoft*
- Wu, Michael, Ph.D*, *Chief AI Strategist at PROS, Instructor at UC Berkeley Extension, and Senior Research Fellow at Ecole des Ponts Business School.*

## Child Development Experts

- Kleinknecht, Erica, Ph.D.*, *Professor, Pacific University Oregon*
- Lulu, Polina*, *Children's Learning Experience Researcher, Designer, at Childrenlx.com*
- Amin Marei, PhD*, *Senior Director, Impact & Efficacy, Sesame Workshop*
- Struttin-Belinoff, Pilar, *Product Design Leader, Sr. Director at Udemy*
- VanderBorght, Mieke, Ph.D*, *Founder of The Global Digital Mindful Project*

**Experts who also contributed to reviewing the report*

## Additional Reviewers

- Pauliac-Vaujour, Emmanuelle, Ph.D, *Attachée for Science and Technology at the Consulate General of France in San Francisco, Embassy of France in the United States of America*
- Penot, Mathieu, *M.Sc in Education at Stanford, Product Manager, IXL Learning*

## Acknowledgments

- Blumberg, Fran, Ph.D, Professor in the Division of Psychological & Educational Services in Fordham University's Graduate School of Education.
- Bornstein, Andrea, MBA from Boston University
- Malvoisin, Celine, Chief Learning Officer at everyone.AI
- Rai, Ajay, Sitrova Innovations
- Seret, Anne-Sophie, Executive Director at everyone.AI
- Spell, Dave, Sitrova Innovations

We extend our sincerest gratitude to the Consulate General of France in San Francisco for their support. Their insights and guidance in forecasting the impact of artificial intelligence on future generations has been instrumental in the production of this report.